\documentclass[eqsecnum,pre,showpacs,twocolumn,a4paper]{revtex4}
\usepackage[english]{babel}
\usepackage{graphicx}
\usepackage{amsmath}
\usepackage{bm}
\usepackage{amsfonts}
\usepackage{amssymb}
\usepackage{color}%
\begin{document}
\title{Finite-size scaling of directed percolation in the steady state}
\author{Hans-Karl Janssen}
\email{janssen@thphy.uni-duesseldorf.de}
\affiliation{Institut f\"ur Theoretische Physik III, Heinrich-Heine-Universit\"at, 40225
D\"usseldorf, Germany}
\author{Sven L\"ubeck}
\email{sven@thp.uni-duisburg.de}
\affiliation{Fachbereich Physik, Universit\"at-Duisburg-Essen, Campus Duisburg,~47048 Duisburg, Germany}
\affiliation{Institut f\"ur Theoretische Physik (B), RWTH Aachen, 52056 Aachen, Germany}
\author{Olaf Stenull}
\email{olaf.stenull@uni-duisburg-essen.de}
\affiliation{Fachbereich Physik, Universit\"at-Duisburg-Essen, Campus Duisburg,~47048 Duisburg, Germany}
\date{\today}

\begin{abstract}
Recently, considerable progress has been made in understanding finite-size
scaling in equilibrium systems. Here, we study finite-size scaling in
non-equilibrium systems at the instance of directed percolation (DP), which has
become the paradigm of non-equilibrium phase transitions into absorbing
states, above, at and below the upper critical dimension. We investigate the finite-size scaling behavior of DP analytically and numerically by considering its steady state
generated by a homogeneous constant external source on a $d$-dimensional
hypercube of finite edge length $L$ with periodic boundary conditions near the
bulk critical point. In particular, we study the order parameter and its
higher moments using renormalized field theory. We derive finite-size scaling
forms of the moments in a one-loop calculation. Moreover, we introduce and calculate a ratio of the order parameter moments that plays a similar role in the analysis of finite size scaling in absorbing nonequilibrium processes as the famous Binder cumulant in equilibrium systems and that, in particular, provides a new signature of the DP universality class. To complement our analytical work, we perform Monte Carlo simulations which confirm our analytical results.
\end{abstract}
\maketitle



\section{Introduction}

\label{intro} \noindent

Critical phenomena like second order phase transitions are characterized by
singularities of various quantities at the
transition point (e.g.~the specific heat, susceptibility, correlation length).
These singularities are described by power-laws governed by critical
exponents. Studying the phase transition of a given system, one usually tries
to identify the set of critical exponents which in conjunction with certain
universal scaling functions characterizes the present universality class.
Powerful analytical and numerical techniques have been developed to accomplish this task.
Analytical investigations of universal quantities allow to address infinite
system sizes but they are usually feasible only if one uses involved
approximations such as the diagrammatic perturbation expansions of
renormalized field theory. Using numerical techniques like Monte Carlo
simulations or transfer matrices calculations one can avoid such
approximations, however, the data is limited to finite systems sizes.
Therefore, finite-size scaling (FSS) is widely used to extrapolate to the
behavior of infinite systems. In particular, FSS is an efficient method to
determine critical exponents and certain universal scaling functions, and
therefore, it often allows to identify the universality class (see
Refs.~\cite{Ba84,Ca88} for reviews). According to the phenomenological FSS
theory~\cite{FiBa72}, finite system sizes $L$ result in a rounding and
shifting of the critical singularities. It is assumed that finite-size effects in isotropic systems 
are controlled sufficiently close to the critical point by the ratio
$L/\xi_{\infty}$, where $\xi_{\infty}$ is the spatial correlation length of
the infinite system. Approaching the transition point, this correlation length
diverges as $\xi_{\infty}\propto r^{-\nu}$,
where $r \propto \left\vert \tau-\tau_{c}\right\vert$ measures the deviation of a temperature-like control parameter $\tau$ from its critical point
value $\tau_{c}$, and where $\nu$ is the critical exponent of $\xi_{\infty}$.
Finite-size effects decrease with increasing $L$ and are negligible for
$L\gg\xi_{\infty}$, i.e., for $L^{1/\nu} r
\gg1$, in systems with periodic boundary conditions, true short range interactions, and without Goldstone modes. Otherwise, they are relevant, i.e., rounding and shifting effects occur
when $L\lesssim\xi_{\infty}$. It is well known that in equilibrium the
hypothesis of the fundamental role of the ratio $L/\xi_{\infty}$ is valid only
below the so-called upper critical dimension $d_{\mathrm{c}}$
(see~\cite{BrDaTo00} for a recent review). Above $d_{\mathrm{c}}$, mean field
theories provide exact results for the critical exponents and the scaling
functions. However, usual FSS fails above $d_{\mathrm{c}}$ because certain
parameters, which are irrelevant in the sense of the renormalization group,
become dangerously irrelevant for $d>d_{\mathrm{c}}$~\cite{Fi74}. Dangerous
irrelevant parameters affect the scaling behavior qualitatively and
furthermore cause the breakdown of hyperscaling laws which connect the
critical exponents to the spatial dimensions~$d$. Investigations of this
breakdown of usual finite scaling date back to the work of Brezin and
Zinn-Justin~\cite{Br82, BrZi85, ZJ96}. For the case of periodic boundary
conditions, Brezin and Zinn-Justin introduced an analytic technique which
makes it possible to perform calculations of size dependent universal scaling
functions. This method exploits the fact that the so-called lowest or zero
mode is distinguished in the sense that in perturbation theory it becomes
critical before the higher modes do and that, therefore, the latter modes can
be traced out perturbatively and fully neglected above $d_{\mathrm{c}}$.

It must be emphasized that a meaningful, quasi-universal analytical study of finite-size effects is possible only in a regime where $1/L \ll 1$ and $r \ll 1$, where it is understood that $L$ and $r$ are measured in terms of suitable non-universal amplitudes. Outside this regime, in particular, if $L$ becomes smaller, finite-size effects will be blurred by the effects of variables that are irrelevant with respect to the corresponding bulk universality class. For $L=O(1)$, analytic approaches are ultimately hopeless. Above $d_{\mathrm{c}}$, the strongest irrelevant effects stem from the usual coupling constant (in the following denoted $g$) of the non-harmonic term in the field theoretic functional, which is relevant below $d_{\mathrm{c}}$, but which is dangerously irrelevant above $d_{\mathrm{c}}$.

\begin{figure}[t]
\includegraphics[width=7cm]{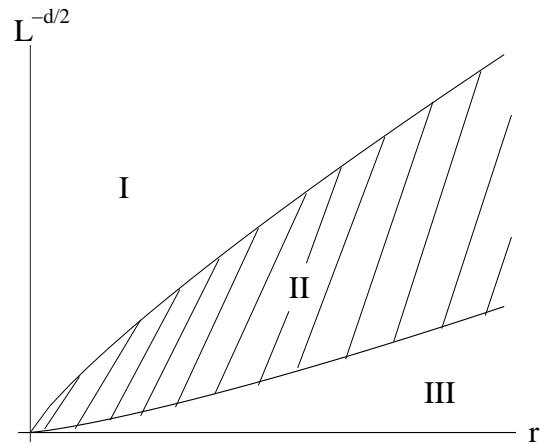}\caption{Scaling regions (schematically) above $d_c$
where corrections to the lowest mode approximation resulting from higher modes
are essential (I and III) or negligible (II).}%
\label{Bereiche}%
\end{figure}

After controversial discussions of the zero-mode theory and the influence of
the higher modes (see e.g.\ \cite{BrDaTo00,Luietal,CheDo98} and references
therein) the problem was recently resolved by Chen and Dohm~\cite{CheDo98},
and convincing agreement between numerical data and field theoretical results
was achieved \cite{St00}. Chen and Dohm showed that even above the upper
critical dimension $d_{\mathrm{c}}$ the higher modes play an essential role.
The following three points summarize key findings: (i) The
higher modes induce a shift of the critical value of the control parameter
proportional to $g^{2}L^{2-d}$, where $g$ is the dangerously irrelevant
coupling constant, cf.\ region I in Fig.~\ref{Bereiche}. This shift is crucial
for the correct interpretation of simulations. (ii) The influence of the
higher modes is essential for the correct description of the exponential
decrease of the finite size effects approaching the infinite volume limit,
cf.\ region III in Fig.~\ref{Bereiche}. Points (i) and (ii) suggest that the
corrections induced by the higher modes can be neglected only in the region
\begin{equation}
L^{d/2-2}\gg L^{d/2}r\gg g^{2}L^{2-d/2}\,, \label{Ber}%
\end{equation}
cf.\ region II in Fig.~\ref{Bereiche}~\cite{regions}.
(iii) Chen and Dohm shed light on the fact that analytical methods using a
hard momentum cutoff, which is well known to be equivalent to long range
interactions, induce a wrong algebraic decrease of finite size
effects. Hence, the widely used Fisher-Wilson momentum shell, like any other
hard-cutoff renormalization procedure, is incompatible with the exponentially
decreasing crossover to the infinite volume limit.

Compared to the equilibrium situation, much less is known in the case of non-equilibrium phase transitions. This motivates us to discuss in the paper at hand FSS in non-equlibrium phase transitions at the
instance of directed percolation (DP). Due to its robustness and ubiquity
(including critical phenomena in physics, biology, dynamics of populations,
epidemiology, as well as autocatalytic chemical reactions) DP is recognized as
the paradigm of non-equilibrium phase transitions into absorbing states
(see~\cite{Hi00,Od04,Lu04} for a recent review on absorbing state transitions, and \cite{JaTa05} for a recent review on renormalized field theory applied to
percolation processes) and, although an exact analytical solution is still
lacking, DP plays a role for non-equilibrium phase transitions comparable to
that of the Ising model for equilibria. Previous studies of finite-size
scaling of percolation processes by one of us and coworkers \cite{JSS88}
focused on the absorbing phase below $d_{\mathrm{c}}$. Here our scope is
different: we are interested in finite-size properties of the steady state below,
at and above $d_{\mathrm{c}}=4$. To be specific, we study for these dimensions
the scaling behavior of finite DP systems in the active phase
which is maintained by a homogeneous external source. Using Reggeon field
theory (RFT) \cite{Gri68}, the generic field theoretic description of the DP universality
class \cite{GraSu78,Ob80,CaSu80,Ja81,Ja01}, we derive finite-size scaling
exponents and universal scaling functions for periodic boundary conditions. For $d>4$, we
demonstrate that the usual phenomenological FSS theory for DP has to be
modified, analogous to what we have discussed above for the equilibrium case,
in order to describe the scaling behavior within the mean field regime. We
show that the correct scaling variable in the strong finite size
region $L\ll\xi_{\infty}\propto r^{-1/2}$ is proportional to $L^{d/2}r,$ and
that corrections, which are controlled by an expansion in a variable $v\propto
gL^{2-d/2}$, become essential only if this variable goes to zero. 

Compared to the equilibrium case, an additional conceptual problem arises in dynamics: to
obtain analytical results for the finite-size scaling functions, one is
forced to perform a Markovian approximation of the dynamics of the lowest
mode. Therefore, our analytical results are restricted to the strong
finite-size region. Outside this region the Markovian approximation leads to a
description of the crossover to the infinite-volume limit by algebraically
decreasing correction terms instead of the correct exponentially decreasing
ones, even if we include the one-loop corrections arising from the higher
modes. We explicitly demonstrate this failure of the Markovian approximation
via a perturbation calculation of the correlation function. 

In the strong finite-size region, we observe, when in the region near the critical point the
shift induced by the higher modes is taken into account, convincing
quantitative agreement between the lowest mode finite-size analysis and our
numerical results. For $d<4$ and $d=4$ we calculate $\varepsilon$-expansions and logarithmic corrections, respectively, for various quantities, focussing, in particular, on an universal ratio of order parameter-moments. For $d<4$, in addition, we perform simulations which clearly underscore that this ratio is a universal signature of the DP class. Brief account of parts of the work presented here has been given previously in Ref.~\cite{LuJa05}.

The outline of our paper is as follows. In Sec.~II we briefly review RFT as
the field theoretic model of choice for the DP universality class. We
derive the effective response functional, i.e., the dynamic free energy of the
homogeneous (lowest) mode. In Sec.~III we calculate this dynamic free energy
in a $1$-loop Markovian approximation. In section~IV we derive finite-size
scaling forms for spatial dimensions above the upper critical dimension. The
steady state solution of the Fokker-Planck equation which correspondents to
the effective response functional yields all moments of the homogeneous mode
in scaling form. In Sec.~V, we compare our analytical results with numerical
results stemming from our Monte Carlo simulations. In Sec.~VI we study the
crossover to mean field theory in the infinite volume limit. In Sec.~VII we
consider finite size effects in the steady state for spatial dimensions below
the upper critical dimension. We apply the renormalization procedure to our
$1$-loop results, and we derive universal values of the aforementioned ratio of order-parameter moments in an $\varepsilon$-expansion. The analytic estimates which follow from this
expansion are compared with the numerical results. In Sec.~\ref{atDc}, we study finite size effects right at the upper critical dimension. We calculate logarithmic corrections to various quantities including our momenta ratio. Concluding remarks are given in Sec.~\ref{conclusions}. An appendix contains a brief presentation of the properties of some functions fundamental to finite-size scaling in DP. For the convenience of the reader, we will provide at the beginning of the main sections short summaries of their respective contents and we point out to their most important formulas. 

\section{Reggeon field theory and the effective response functional}
\label{RFT} 

We start our analysis by deriving an effective response functional for the zero mode. One of the main findings of this section is that the distance $r$ from the bulk critical point in this effective theory is given by Eq.~(\ref{rDef}). Equation~(\ref{freie energie}) summarizes our result for the effective response functional.

It has been known for a long time that the DP universality class is well
represented by RFT. For a recent overview on the field theories of percolation
processes and the derivation of the underlying minimal models from basic
principles see \cite{JaTa05}. RFT, based originally upon a non-hermitean
Hamilton-operator \cite{Gri68}, is equivalent to a Langevin-description of a
minimal DP process, the so-called Gribov process \cite{GraSu78}. After reduction to the
relevant terms, the stochastic equation of motion of this DP process may be
written in the form of the Langevin equation (in the It\^{o} interpretation)
\cite{Ja81,Ja01}:
\begin{equation}
{\lambda}^{-1}\partial_{t}s(\mathbf{r},t)=-\left[  \nabla^{2}+\tau+\frac{g}%
{2}s(\mathbf{r},t)\right]  s(\mathbf{r},t)+h+\zeta(\mathbf{r},t).
\label{eq:langevin_dp_01}%
\end{equation}
Here, the activity field $s(\mathbf{r},t)\geq 0$, which is proportional to the
density of active particles (agents) on a mesoscopic (coarse grained) scale,
is the order parameter field of the non-equilibrium phase transition. The diffusional term represents the isotropic spreading of activity. The
control parameter of the transition is $\tau$, and $\tau_{c}$ denotes its
critical value. In the infinite volume limit, a finite positive particle
density occurs below the transition point ($\tau<\tau_{c}$) whereas the
absorbing vacuum state ($s=0$) is approached above the transition point if the
source term $h\geq 0$ (which can be implemented in simulations, e.g., as a
spontaneous particle creation process~\cite{LuWi02}) is absent. In a finite
system, the absorbing state is inevitably approached even for $\tau<\tau_{c}$,
if $h=0$. However, it can be shown \cite{Ja05} that the logarithm of the
relaxation time to the absorbing state increases proportional to the system
volume in the active phase below $\tau_{c}$. $\lambda$ and $g$ denote the
kinetic and coupling constants, respectively. $\zeta$, finally, represents the
noise which accounts for fluctuations of the particle density. All universal
properties of the DP universality class are captured by the minimal model,
Eq.~(\ref{eq:langevin_dp_01}), provided the noise $\zeta(\mathbf{r},t)$ is a
Gaussian random variable with zero mean and correlator given
by~\cite{footnoteG}
\begin{equation}
\overline{\zeta(\mathbf{r},t)\zeta(\mathbf{r}^{\prime},t^{\prime})}%
=\lambda^{-1}g'\,s(\mathbf{r},t)\delta(\mathbf{r}-\mathbf{r}^{\prime}%
)\delta(t-t^{\prime}). \label{eq:langevin_dp_corr_01}%
\end{equation}
Note, that only an absorbing noise with a correlator that comprises at least a
term linear in the field $s$ ensures that the systems is trapped
in the absorbing state with a continuously decreasing survival probability
\cite{Ja05}. A form of the noise proportional to $s^{2}$ (multiplicative
noise) results in a survival probability which is strictly $1$ for all finite times.

Renormalization group techniques have been applied quite successfully to
determine the critical exponents and universal scaling functions of
DP~\cite{Ja81,Ja01,Ja05,JaKuOe99,JaTa05,CaSu80,Ob80,JaSt04}. In the framework of
field theory, a path integral formulation of stochastic processes is more
useful than their Langevin equations. In the path integral formulation,
correlation functions and response functions can be determined by calculating
path integrals with weight $\exp{(-\mathcal{J})}$ \cite{Ja76,DeDo76}, where
the dynamic response functional $\mathcal{J}$ describes the considered
stochastic process. The dynamic response functional of the Gribov
process~(\ref{eq:langevin_dp_01}) is given by~\cite{Ja81,Ja01}
\begin{equation}
\mathcal{J}[\tilde{s},s]=\int\mathrm{d}^{d}{r}\,\mathrm{d}t\,{\lambda
}\Big\{{\tilde{s}}\Big[\lambda^{-1}\partial_{t}+\bigl(\tau-\nabla
^{2}\bigr)+\frac{g}{2}\bigl({s}-\tilde{s}\bigr)\Big]s-h{\tilde{s}}\Big\} ,
\label{eq:action_reggeon_field_theory}%
\end{equation}
where ${\tilde{s}}(\mathbf{r,}t)$ denotes the purely imaginary response field conjugated to the
Langevin noise field. The functional $\mathcal{J}$ is invariant under time
reversal (in RFT usually called rapidity reversal)
\begin{equation}
{\tilde{s}}(\mathbf{r},t)\,\longleftrightarrow\,-\,{s}(\mathbf{r},-t)\,
\label{eq:rapidity_trans}%
\end{equation}
as long as the (symmetry breaking) field $h$ vanishes. This symmetry is
spontaneously broken in the active phase below the transition point. In
general, the time reversal invariance of the minimal model is merely an
asymptotic symmetry of systems belonging to the DP universality class. Note,
however, that this symmetry is exact for bond DP.

It is worth noting, that the original RFT \cite{Gri68} is based on a bosonic
annihilation-creation formalism in which $s$ is related to the annihilation
operator, and $\tilde{s}$ to the creation operator (for a recent review over this master-equation approach see \cite{TaHoVL05}). Hence, as
described in \cite{JaTa05}, the original RFT and the fluctuating field theory
based on the Gribov process (\ref{eq:langevin_dp_01}), where $s$ is
proportional to a real positive density, are formally different. Note that the bosonic theory leads to an additional noise term proportional to $(\tilde{s}s)^2$ in the functional $\mathcal{J}$, Eq.~(\ref{eq:action_reggeon_field_theory}), with positive sign. Such noise terms result from anticorrelating more-particle annihilation reactions, and are typical for diffusion-limited reactions. However, for DP, which is not a diffusion-limited reaction system, this noise term is irrelevant. Hence, both formalism, the Langevin- and the master-equation approach, produce the same perturbation series which leads via the renormalization group to the same universal asymptotic behavior. Note, however, that in the bosonic formalism $s$ and $\tilde{s}$ are constructed as complex fields with $\tilde{s}s$ real and positiv. Thus, after deleting the irrelevant noise term, the functional integration with the weight $\exp{(-\mathcal{J})}$ is a priori mathematically ill-defined. Ciafaloni and Onofri \cite{CiaOn79} have shown more than 25 years ago that in this case the only correct support for integration over $s$ and $\tilde{s}$ is, respectively, the real positive axis and the full imaginary axis (see also the appendix of \cite{Ja01} for a corresponding quasi-canonical transformation of the fields). The upshot is that only the Langevin equation formalism offers a mathematically correct interpretation of the functional integral. Because we must use at least parts of the weight $\exp{(-\mathcal{J})}$ as it stands, without applying perturbation theory, as a probability measure, this interpretation is of greatest importance.

Using standard techniques known from equilibrium~\cite{Br82,BrZi85,ZJ96}, one
can extract from $\mathcal{J}$ an effective response functional for the zero
mode, which then can be used to calculate size-dependent universal scaling
functions as well as the involved critical exponents. To follow this route,
let us consider DP in a finite cubic geometry of linear size~$L$ with periodic
boundary conditions and expand $s$ and $\tilde{s}$ in plane waves,
\begin{equation}
s(\mathbf{r},t)=\sum_{\mathbf{q}}\mathrm{e}^{i\mathbf{q\cdot r}}%
s(\mathbf{q},t)\, , \label{eq:plane_waves}%
\end{equation}
and likewise for $\tilde{s}$. Each component of the wavevector $\mathbf{q}$
takes on discrete values, viz.\ multiples of $2\pi/L$ including zero. When
dealing with summations over $\mathbf{q}$, one has to bear in mind that
path-integrals based on the response
functional~(\ref{eq:action_reggeon_field_theory}) are well-defined only if an
appropriate regularization of the diverging UV behavior is applied. In
principle, there are different options for choosing a regularization
procedure. As discussed in the introduction and as can be easily checked by applying the Euler-McLaurin summation formula, a hard momentum cutoff (support
of the modes only for momenta with $\left\vert \mathbf{q}\right\vert
\leq\Lambda$) is inappropriate for studying FSS, since a hard cutoff induces
non-physical long-range correlations in real space \cite{ZJ96} which
contaminate the finite size calculations \cite{CheDo98}. Lattice
regularization, where the system is placed on a discrete lattice instead of
spatial continuum, is the most physical one. Moreover, this regularization is
in closest contact to simulations. However, lattice-regularization replaces
the Laplacian by the lattice difference-operator. Thus, analytical
calculations become very complicated. In the following, we will use
(implicitly) a soft cutoff procedure, i.e., we will include a factor
$\exp(-\mathbf{q}^{2}\Lambda^{2})$ in all summations over wave vectors
$\mathbf{q}$, followed by dimensional regularization and the limit
$\Lambda\rightarrow\infty$. One can show that this procedure is equivalent in
the scaling region to lattice regularization as long as one concentrates on
universal quantities. At this point, a word of caution is in order. If very
small lattices are considered, it may be more appropriate to use lattice
regularization~\cite{CheDo98}. In the following, we will ignore very small
lattices in our analytic considerations because for these lattices one has to
expect many non-universal corrections.

The Fourier transformed propagator of the perturbation theory about the
saddle-point of the path-integrals (mean-field theory) based on the response
functional Eq.~(\ref{eq:action_reggeon_field_theory}) is given by
\begin{equation}
G_{\mathrm{0}}(\mathbf{q},\omega)=\frac{1}{i\omega/\lambda+r+\mathbf{q}^{2}%
}\,, \label{G_full}%
\end{equation}
with
\begin{equation}
\label{rDef}
r=(\tau-\tau_{\mathrm{c}})+M
\end{equation}
measuring the distance to the critical point. Here
$M=g\langle s\rangle$, with the expectation value $\langle s\rangle$
determined by the condition that tadpoles are excluded in the
diagrammatical perturbation expansion. In mean-field theory, $r=\sqrt{\tau
^{2}+2gh}$ and $\tau_{c}=0$. Hence, for small frequencies
$\omega$ in the finite-size limit, $w=r(L/2\pi)^{2}\ll1$, the zero mode with
$\mathbf{q}=0$ separates from the higher modes and leads to infrared
divergencies in perturbation theory. Therefore, functional integrals of the
zero mode must be calculated exactly, and cannot be handled by perturbation
theory~\cite{Br82,BrZi85,ZJ96}. Perturbation theory can be used, however, as a
tool for the functional integration of the higher modes. As we will discuss in
detail later on, the Gaussian fluctuations of the higher modes have a
significant influence on the scaling functions describing the crossover from
$w\approx1$ to $w\gg1$, as well as the behavior near the bulk critical point
$w\approx0$. Nonetheless, mean-field theory should be correct for $d>4$ in the
bulk limit $w\rightarrow\infty$.

Following \cite{JSS88}, we construct an effective response functional for the
zero-mode by separating the homogenous mode $\Phi(t)$ from its orthogonal
complements $\Psi(\mathbf{r},t)$ via setting
\begin{equation}
g\,s(\mathbf{r},t)=\Phi(t)+\Psi(\mathbf{r},t) \label{eq:decomposing}%
\end{equation}
with $\Phi(t)=gL^{-d}\int\mathrm{d}^{d}r\,s(\mathbf{r},t)$ and likewise for $\tilde{s}$.  This leads to a
decomposition of the action, ${}\mathcal{J}=\mathcal{J}_{0}+\mathcal{J}%
_{1}+\mathcal{J}_{2}$, with
\begin{equation}
\mathcal{J}_{0}=\lambda g^{-2}L^{d}\int\mathrm{d}t\,\Big\{\tilde{\Phi
}\Big[\lambda^{-1}\partial_{t}+\tau+\frac{1}{2}(\Phi-{\tilde{\Phi})}%
\Big]\Phi-H{\tilde{\Phi}}\Big\}\,, \label{J_PHI}%
\end{equation}
where $H=gh$, and
\begin{subequations}
\label{Jot1,2}%
\begin{align}
\mathcal{J}_{1}  = & \,  \lambda g^{-2}\int\mathrm{d}^{d}{r}\,\mathrm{d}
t\,\Big\{\tilde{\Psi}\Big[\lambda^{-1}\partial_{t}+\bigl(\tau-\nabla
^{2}\bigr)+(\Phi-{\tilde{\Phi})}\Big]\Psi
\nonumber \\
&\qquad\qquad\qquad\qquad +\frac{1}{2}\bigl(\tilde{\Phi}%
\Psi^{2}-\Phi\tilde{\Psi}^{2}\bigr)\Big\},\label{J_1}\\
\mathcal{J}_{2}  = &\, \lambda\frac{g^{-2}}{2}\int\mathrm{d}^{d}{r}%
\,\mathrm{d}t\,\tilde{\Psi}(\Psi-\tilde{\Psi}{)\Psi.} \label{J_2}%
\end{align}
We have included the coupling constant $g$ in the definition of the fields
$\Phi$, $\tilde{\Phi}$, $\Psi$, $\tilde{\Psi}$ to disentangle the two
different roles of $g$, which on the one hand serves as the loop-order
generating parameter of the perturbation theory around the mean-field
(Landau) approximation, and on the other hand is a scale factor of the fields. This
last role is what makes $g$ a \textquotedblleft dangerous\textquotedblright%
\ irrelevant variable, as alluded to in the introduction. Finally, we
eliminate $\tilde{\Psi}$ and $\Psi$ via functional integration. This leads to
\end{subequations}
\begin{equation}
\Sigma\lbrack{\tilde{\Phi},}\Phi]=\mathcal{J}_{0}-\ln\int\mathcal{D[}%
\tilde{\Psi},{\Psi]\exp}\bigl(-\mathcal{J}_{1}-\mathcal{J}_{2}%
\bigr) \label{freie energie}%
\end{equation}
as our effective response functional for the homogeneous mode. In the
following, we will also refer to $\Sigma$ as our dynamic free energy.

The zero-loop approximation $\Sigma\approx\mathcal{J}_{0}$ is known as the
lowest mode approximation of finite-size scaling \cite{BrZi85,ZJ96}. As we
move along, we will show that, for $d>4$, this lowest mode theory is modified
outside the lowest-mode region $L^{d/2-2}\gg rL^{d/2}\gg L^{2-d/2}$, see
Fig.~(\ref{Bereiche}), by one-loop (Gaussian) contributions arising from the
higher modes.

\section{Dynamic free energy in the one-loop expansion}
\label{Dyn Freie Energie} 

In this section, we calculate the dynamic free energy $\Sigma$ to 1-loop order in perturbation theory. Key formulas of this section are Eqs.~(\ref{Def_w}) and (\ref{Def_v}) which give the finite-size scaled version of, respectively, the distance $r$ from the bulk critical point and the dangerously irrelevant coupling constant $g$. Central to our  discussions to follow is the 1-loop dynamical free energy~(\ref{vollst_freie_energie}) in conjunction with our 1-loop results for the parameters appearing in it, Eq.~(\ref{Par2}).

To 1-loop order, $\mathcal{J}_{2}$ does not contribute and hence can be neglected.
$\mathcal{J}_{1}$ contributes via the propagator
\begin{align}
G(t,t^{\prime};\mathbf{q})= & \; \theta(t-t^{\prime})\,\exp\bigg[-\lambda
(\tau+\mathbf{q}^{2})(t-t^{\prime})
\nonumber \\
&\qquad +\lambda\int_{t^{\prime}}^{t}
dt^{\prime\prime}\bigl(\tilde{\Phi}(t^{\prime\prime})-\Phi(t^{\prime\prime
})\bigr)\bigg]. \label{Prop}
\end{align}
of the higher modes, which is determined by the bilinear part in the fields
$\tilde{\Psi}$, ${\Psi}$ of $\mathcal{J}_{1}$. Gaussian integration yields
readily
\begin{align}
-\ln\int\!\!\mathcal{D[}\tilde{\Psi},{\Psi]}&\exp\bigl(-\mathcal{J}_{1}
\bigr) =\frac{\lambda^{2}}{2}\sum_{\mathbf{q}\neq\mathbf{0}}
{\displaystyle\iint}
dt\,dt^{\prime}\,\tilde{\Phi}(t)
\nonumber \\
&\times G(t,t^{\prime};\mathbf{q})^{2}\Phi(t^{\prime
})+O((\tilde{\Phi}\Phi)^{2})\,. \label{Gauss_Int}%
\end{align}
For the time being, let us concentrate on the region $w\ll1$. Then, the
typical time-dependence of the zero-mode shows slowing down in comparison to
the higher modes leading to Markovian behavior of the zero-mode. Thus, we can
approximate $\Phi(t^{\prime})$ in Eq.~(\ref{Gauss_Int}) by $\Phi
(t)-(t-t^{\prime})\dot{\Phi}(t)$ and the propagator simplifies to
\begin{equation}
G(t,t^{\prime};\mathbf{q})=\theta(t-t^{\prime})\,\exp\Big[-\lambda
\bigl(\tau+\mathbf{q}^{2}+\Phi(t)-\tilde{\Phi}(t)\bigr)(t-t^{\prime})\Big]\,.
\label{Prop_appr}%
\end{equation}
Note that this Markovian approximation does not any longer allow a correct
description of the crossover from the finite size to the infinite volume
behavior. If one incorrectly takes $w\gg1$ in the results following from this
approximation one gets algebraically decreasing correction terms describing
the crossover to the infinite volume limit. This crossover is qualitatively
wrong because the corrections must be exponentially decreasing. We will
discuss this shortcoming of the Markovian approximation in
Sec.~\ref{crossoverToMeanField}, where we calculate the steady state
correlation function for $w\gg1$ in a $1$-loop calculation.

After application of the Markovian approximation the residual time integration
of $t^{\prime}$ can be done. We obtain%
\begin{align}
&-\ln\int\mathcal{D[}\tilde{\Psi},{\Psi]\exp}\bigl(-\mathcal{J}_{1}\bigr)
\nonumber \\
&=\int
dt\sum_{\mathbf{q}\neq\mathbf{0}}\bigg\{\frac{\lambda\tilde{\Phi}\Phi
}{4\bigl(\tau+\Phi-\tilde{\Phi}+\mathbf{q}^{2}\bigr)}-\frac{\tilde{\Phi}%
\dot{\Phi}}{8\bigl(\tau+\Phi-\tilde{\Phi}+\mathbf{q}^{2}\bigr)^{2}}\biggr\}\,
\label{Res_FunkInt}%
\end{align}
retaining only terms of the form already appearing in $\mathcal{J}_{0}$, i.e.,
neglecting fourth-order terms in $\Phi$ and $\tilde{\Phi}$. These higher order
monomials lead to corrections of higher order in $L^{-1}$ as the retained ones
\cite{BrZi85,ZJ96}. Subsequently, we expand the denominators in
Eq.~(\ref{Res_FunkInt}) in $\Phi$ and $\tilde{\Phi}$ about their mean values
$\langle\Phi\rangle=M$ and $\langle\tilde{\Phi}\rangle=0$. Note that this
procedure provides strictly positive denominators even in the case
$\tau_{\mathrm{c}}-\tau\geq(2\pi/L)^{2}$ and also that we can include the bulk
critical value $\tau_{\mathrm{c}}$ of the control parameter $\tau$ in the
denominators of Eq.~(\ref{Res_FunkInt}) since $\tau_{\mathrm{c}}$ is of order
$g^{2}$. Recalling definition~(\ref{rDef}),
we finally obtain from
Eqs.~(\ref{J_PHI}) (\ref{freie energie}) and (\ref{Res_FunkInt}) that
\begin{equation}
\Sigma\lbrack{\tilde{\Phi},}\Phi]=\lambda g^{-2}L^{d}\int\mathrm{d}%
t\,\Big\{\tilde{\Phi}\Big[\lambda^{-1}\hat{k}\partial_{t}+\hat{\tau}%
+\frac{\hat{f}}{2}(\Phi-{\tilde{\Phi})}\Big]\Phi-H{\tilde{\Phi}}\Big\}\,.
\label{vollst_freie_energie}%
\end{equation}
The parameters $\hat{k}$, $\hat{\tau}$ and $\hat{f}$ are given by
\begin{subequations}
\label{Par}%
\begin{align}
\hat{k}(r)  &  =\Big[1-\frac{g^{2}}{8}S_{2}(r)\Big]\,,\label{k_dach}\\
\hat{\tau}(r)  &  =\Big[1-\frac{g^{2}}{4}S_{2}(r)\Big]\tau+\frac{g^{2}}%
{4}\Big[S_{1}(r)+rS_{2}(r)\Big]\,,\label{tau_dach}\\
\hat{f}(r)  &  =\Big[1-\frac{g^{2}}{2}S_{2}(r)\Big]\,, \label{f_dach}%
\end{align}
with $S_{l}$ defined by
\end{subequations}
\begin{equation}
S_{l}(r)=L^{-d}\sum_{\mathbf{q}\neq\mathbf{0}}\frac{1}{\bigl(r+\mathbf{q}%
^{2}\bigr)^{l}}\,. \label{S_l}%
\end{equation}
As mentioned earlier, all sums over wavevectors must be regularized
appropriately; it is understood that in actual calculations these sums are
augmented by a soft cutoff factor $\exp(-\mathbf{q}^{2}\Lambda^{2})$. In the
infinite-size limit, $L\rightarrow\infty$, the sums $S_{l}(r)$ tend to the
integrals
\begin{equation}
S_{l}^{\infty}(r)=\int_{\mathbf{q}}\frac{1}{\bigl(r+\mathbf{q}^{2}\bigr)^{l}%
}\, \label{S_l_inf}%
\end{equation}
with $\int_{\mathbf{q}}\ldots=(2\pi)^{-d}\int d^{d}q\ldots$. The bulk critical
point is then given in $1$-loop approximation by%
\begin{equation}
\tau_{\mathrm{c}}=-\frac{g^{2}}{4}S_{1}^{\infty}(0)=-\frac{g^{2}}{4}%
\int_{\mathbf{q}}\frac{1}{\mathbf{q}^{2}}\,. \label{tau_c}%
\end{equation}

For the steps to follow, it is useful to introduce the differences
\begin{equation}
\Delta_{l}(r)=S_{l}^{\infty}(r)-S_{l}(r)=:\frac{L^{2l-d}}{(2\pi)^{2l}%
\Gamma(l)}\,D^{(l)}(w) \label{Def_D}%
\end{equation}
with the scaling variable
\begin{equation}
w=\Big(\frac{L}{2\pi}\Big)^{2}r\,. \label{Def_w}%
\end{equation}
The $D^{(l)}$ are functions of this scaling variable given by
\begin{align}
D^{(l)}(w)&=\int_{0}^{\infty}dt\,t^{l-1}\mathrm{e}^{-wt}\,\Big[\Big(\frac{\pi
}{t}\Big)^{d/2}-A(t)^{d}+1\Big]
\nonumber \\
&=-\partial_{w}D^{(l-1)}(w) \label{D_l}%
\end{align}
where $A(t)=1+2\sum_{n=1}^{\infty}\exp(-n^{2}t)=(\pi/t)^{1/2}A(\pi^{2}/t)$.
Some important properties of the functions $D^{(l)}(w)$ are discussed in the
appendix. With help of the differences~(\ref{Def_D}), we can express the
parameters appearing in the dynamic free energy after some rearrangements as,
\begin{subequations}
\label{Par1}%
\begin{align}
\hat{k}(r)  &  =\Big[1-\frac{g^{2}}{8}\int_{\mathbf{q}}\frac{1}{(r+\mathbf{q}%
^{2})^{2}}\Big]+\frac{g^{2}}{8}\Delta_{2}(r)\,,\label{k_dach_re}
\\
\hat{\tau}(r)  &  =\Big[1-\frac{g^{2}}{4}\int_{\mathbf{q}}\frac{1}
{(r+\mathbf{q}^{2})^{2}}\Big]\,(\tau-\tau_{\mathrm{c}})
\nonumber \\
&-\frac{g^{2}}{4}
\int_{\mathbf{q}}\frac{r^{2}}{\mathbf{q}^{2}(r+\mathbf{q}^{2})^{2}}\nonumber\\
& +\frac{g^{2}}{4}\bigl[\Delta_{2}(r)\,(\tau-\tau_{\mathrm{c}}%
)-\Delta_{1}(r)-r\Delta_{2}(r)\bigr]\,,\label{tau_dach_re}
\\
\hat{f}(r)  &  =\Big[1-\frac{g^{2}}{2}\int_{\mathbf{q}}\frac{1}{(r+\mathbf{q}%
^{2})^{2}}\Big]+\frac{g^{2}}{2}\Delta_{2}(r)\,, \label{f_dach_re}%
\end{align}
where we have neglected terms of order $g^{4}$. The integrals over wavevectors
in Eqs.~(\ref{Par1}) lead to IR singularities for spatial dimensions $d\leq4$
if $r\rightarrow0$. These singularities must be treated by the renormalization
group \cite{JSS88}. We will return to the cases $d<4$ and $d=4$ in Secs.~\ref{belowDc} and \ref{atDc}, respectively.

For $d>4$, the integrals lead to cut-off dependent nonuniversal constants up
to corrections of order $r^{(d-4)/2}$. We neglect these corrections, and
include the nonuniversal constants in a rescaling of the fields $\Phi$,$\tilde{\Phi}$,
and of the parameters $\tau$,$g$,$H$. We redefine $\tau
-\tau_{\mathrm{c}}\rightarrow\tau$ and thus, henceforth, $\tau=0$ at the bulk
critical point. Finally, we obtain for $d>4$
\end{subequations}
\begin{subequations}
\label{Par2}%
\begin{align}
\hat{k}   = & \Big[1+\frac{v^{2}}{2}D^{(2)}(w)\Big]\,,\label{k_dach_end}\\
\hat{\tau}  = & \Big[1+v^{2}D^{(2)}(w)\Big]\tau \nonumber\\
&-v^{2}\Big[D^{(1)}
(w)+wD^{(2)}(w)\Big]\,\Big(\frac{2\pi}{L}\Big)^{2}\,,\label{k_tau_end}\\
\hat{f}  = & \Big[1+2v^{2}D^{(2)}(w)\Big]\,. \label{f_dach_end}
\end{align}
where we have defined a second scaling variable%
\end{subequations}
\begin{equation}
v=\frac{g}{8\pi^{2}}L^{2-d/2}\,. \label{Def_v}%
\end{equation}

Now, after having identified $w$ and $v$ as fundamental scaling variables, it
is worthwhile to briefly reconsider the condition for the approximations that
we made in this section. To justify the neglect of higher loop-orders of the
perturbation expansion as well as the influence of other irrelevant couplings
in the response functional~(\ref{eq:action_reggeon_field_theory}) we have to
assume $v^{2}\ll1$, that means that $L$ is sufficient large but finite. Moreover, for the application of the Markovian approximation, we have to assume $w \ll 0$.

\section{Fokker-Planck equation and scaling of the steady state observables
above $\bm{d_{\text{c}}}$}

\label{Scaling} \noindent

In this section we analytically derive scaling forms and scaling functions for steady state observables above the upper critical dimension. First, we identify further fundamental scaling variables, namely the finite-size scaled control parameters given in Eq.~(\ref{x_and_y}). Moreover, we introduce finite-size scaled fields, Eq.(\ref{Fi-fi}), and a finite size-scaled time, Eq.~(\ref{t-s}). This leads to a finite-size scaled dynamic free energy, Eq.~(\ref{resc_free_energy}), with parameters $a$ and $b$ given in Eq.~(\ref{a_und_b}), which will play a central role as we move along. Then, we discuss how we can calculate the moments of the homogeneous density, i.e., averages of powers of $\Phi$, with the help of a Fokker-Planck equation, Eq.~(\ref{FP_Eq}), and its stationary solution, Eq.~(\ref{Stat_Vert}). Our results for the moments of the homogeneous density are presented in Eq.~(\ref{Momente}).

As we move along, we shall see that, besides $w$ and $v$, three further
combinations of the four physical parameters $\tau$, $h$, $M$, $L$ and the
dangerous irrelevant coupling constant $g$ emerge as natural variables of the
finite-size scaling forms of the moments of the homogeneous density, namely
\begin{equation}
x=\frac{2}{g}L^{d/2}\tau\,,\qquad y=\frac{2}{g^2}L^{d}H\,,\qquad z=\frac{2}%
{g}L^{d/2}M\,. \label{x_and_y}%
\end{equation}
The variables $w$ and $v$ are related to $x$ and $z$ by
\begin{equation}
w=v\,(x+z)\,. \label{w_und_m}%
\end{equation}
Moreover, we introduce scaled fields $\varphi$, $\tilde{\varphi}$, and a
scaled time $s$,
\begin{subequations}
\label{Ord+Zeit}%
\begin{align}
\Phi(t)  &  =\alpha\varphi(s)\,,\quad\tilde{\Phi}(t)=\alpha\tilde{\varphi
}(s)\,,\label{Fi-fi}\\
\lambda t  &  =\beta s\,, \label{t-s}%
\end{align}
with scale factors $\alpha$, $\beta$ given by
\end{subequations}
\begin{subequations}
\begin{align}
\alpha &  =\hat{k}^{-1/2}\,gL^{-d/2}\,,\label{alpha}\\
\beta &  =2\hat{f}^{-1}\hat{k}^{3/2}\,g^{-1}L^{d/2}\,. \label{beta}%
\end{align}
Incorporating these rescalings into the dynamic free
energy~(\ref{vollst_freie_energie}), we obtain
\end{subequations}
\begin{equation}
\Sigma\lbrack\tilde{\varphi}{,}\varphi]=\int\mathrm{d}s\,\Big\{\tilde{\varphi
}\Big[\partial_{s}+a+(\varphi-\tilde{\varphi}{)}\Big]\varphi-b{\tilde{\varphi
}}\Big\}\,, \label{resc_free_energy}%
\end{equation}
with new parameters
\begin{subequations}
\label{abDef}
\begin{align}
a&=2\hat{k}^{1/2}\hat{g}^{-1}\hat{\tau}L^{d/2}
\\
b&=2\hat{k}\hat{g}^{-1}hL^{d},
\end{align}
\end{subequations}
where $\hat{g}=g\hat{f}$. When
expressed in terms the scaling variables, these new parameters read
\begin{subequations}
\label{a_und_b}%
\begin{align}
a  = &\, \Big[1-\frac{7v^{2}}{4}D^{(2)}(w)+O(v^{4})\Big]
\nonumber \\
&\times \Big\{x-vD^{(1)}
(w)-zv^{2}D^{(2)}(w)+O(v^{3})\Big\}\nonumber\\
= &\,  \Big[1-\frac{3v^{2}}{4}D^{(2)}(w)+O(v^{4})\Big]\,x
\nonumber \\
&-v\Big[D^{(1)}(w)+wD^{(2)}(w)+O(v^{2})\Big]\,,\label{a}\\
b  = &\, \Big[1-\frac{3v^{2}}{2}D^{(2)}(w)+O(v^{4})\Big]\,y\,. \label{be}
\end{align}
Here we have indicated $2$-loop contributions and higher order ones,
resulting from neglected irrelevant couplings in the response functional
(\ref{eq:action_reggeon_field_theory}), by the Landau order symbol. The
rescaled form~(\ref{resc_free_energy}) makes transparent an essential feature
of the dynamic free energy $\Sigma\lbrack\tilde{\varphi}{,}\varphi]$: it
depends only on the two parameters $a$ and $b$. As a straightforward
consequence, all correlation and response functions of the homogeneous density
(which is proportional to $\varphi$) are universal functions of $a$, $b$, and
the scaled time $s$:
\end{subequations}
\begin{equation}
F_{N,\tilde{N}}(\{s_{i}\},a,b)=\left\langle {\displaystyle\prod\limits_{i=1}
^{N}}\varphi(s_{i})\!{\displaystyle\prod\limits_{j=N+1}^{N+\tilde{N}}}
\!\!\!\tilde{\varphi}(s_{j})\right\rangle \,. \label{univFu}%
\end{equation}
The strict lowest-mode approximation neglects all the $1$-loop corrections of
the higher modes, that is it sets $v=0$. Hence, we have $a(v=0)=x$ and
$b(v=0)=y$ in this approximation. Recalling the definitions~(\ref{Ord+Zeit}),
we find that the correlation and response functions
\begin{equation}
G_{N,\tilde{N}}(\{t_{i}\},\tau,h,L,g,\lambda)=\left\langle \prod_{i=1}^{N}\Phi
(t_{i})\!\!\prod_{j=N+1}^{N+\tilde{N}}\!\!\tilde{\Phi}(t_{j})\right\rangle
^{(\mathrm{cum})}
\end{equation}
of the homogeneous density have the finite-size scaling form in the zero-loop approximation
\begin{align}
&G_{N,\tilde{N}}(\{t_{i}\},\tau,h,L,g,\lambda)=(g^{2}/L^{d})^{(N+\tilde{N}%
)/2}
\nonumber \\
&\times F_{N,\tilde{N}}(\{L^{-d/2}g\lambda t_{i}/2\},2L^{d/2}\tau/g,2L^{d}h/g)
\label{low-mod-scal}%
\end{align}
with the universal scaling functions $F_{N,\tilde{N}}$.

Next, we determine the scaling functions $F_{N,0}(\{0\},a,b)$ including their $1$-loop
corrections. Path integrals with weight $\exp(-\Sigma\lbrack\tilde{\varphi}%
{,}\varphi])$ based on the dynamic free energy $\Sigma\lbrack\tilde{\varphi
}{,}\varphi]$, Eq.~(\ref{resc_free_energy}), are equivalent to mean values
taken with a probability $P(\varphi,s|\varphi_{0})d\varphi$ to find the
process in the interval $[\varphi,\varphi+d\varphi]$ at time $s$ if the
process is started at time $0$ with $\varphi_{0}$. The probability density
$P(\varphi,s|\varphi_{0})$ is determined by the Fokker-Planck
equation~\cite{Ja05,JSS88}
\begin{align}
\partial_{s}P(\varphi,s|\varphi_{0})=&\;\partial_{\varphi}\big\{\bigl[(a+\varphi
)\varphi-b\bigr]P(\varphi,s|\varphi_{0})\big\}
\nonumber \\
&\qquad +\partial_{\varphi}^{2}\big\{\varphi P(\varphi,s|\varphi_{0})\big\}\,, \label{FP_Eq}%
\end{align}
with initial condition $P(\varphi,0|\varphi_{0})=\delta(\varphi-\varphi_{0})$.
In the classification scheme of Feller \cite{Fe52,BR60}, the infinite point
$\varphi=\infty$ is a natural boundary, and therefore $P(\infty,s|\varphi
_{0})=0$. The boundary at $\varphi=0$ is a so-called exit boundary,
representing the absorbing state as a growing $\delta$-function, if $b=0$. In
the case $b>0$, this boundary is regular (entrance) if $0<b<1$, and natural
for $b\geq1$. In both cases it is easy to find the steady state distribution:%
\begin{equation}
P_{0}(\varphi)=C\mathrm{\,}\varphi^{b-1}\,\exp\Big(-a\varphi-\frac{\varphi
^{2}}{2}\Big), \label{Stat_Vert}%
\end{equation}
where $C$ is determined by the normalization condition $\int_{0}^{\infty
}d\varphi\,P_{0}(\varphi)=1$. Note that in the limit $b\rightarrow0$ the
normalization constant $C$ goes to zero as a consequence of the absorbing
state. In this case the only normalizable stationary probability density is
$\lim_{b\rightarrow0}P_{0}(\varphi)=\delta(\varphi)$.

Now we fix the scaling variable $z$. Because $a$ and $b$ are the only
parameters that our state distribution depends on, $z$ enters the $1$-loop
correction terms, but it does not appear at zero-loop order. Hence, we can use
here the strict lowest mode approximation $z=2\langle\varphi\rangle_{0}$,
where the mean value is calculated with the steady state
distribution~(\ref{Stat_Vert}) with $a$ and $b$ set equal to the zero-loop
forms $a(v=0)=x$ and $b(v=0)=y$. This leads us to
\begin{equation}
z=2\vartheta_{1}(x,y) \, ,
\label{m}%
\end{equation}
where $\vartheta_{1}$ is a member of the set of functions defined by
$\langle\varphi^{l}\rangle=\vartheta_{l}(a,b)$, i.e.,
\begin{align}
\vartheta_{l}(a,b)&=\frac{\int_{0}^{\infty}d\varphi\,\varphi^{b+l-1}
\,\exp\bigl(-a\varphi-\varphi^{2}/2\bigr)}{\int_{0}^{\infty}d\varphi
\,\varphi^{b-1}\,\exp\bigl(-a\varphi-\varphi^{2}/2\bigr)}\nonumber\\
&=\frac{\Gamma(b+l)\,D_{-b-l}(a)}{\Gamma(b)\,D_{-b}(a)}\,. \label{theta_l}
\end{align}
Here, $D_{\alpha}(z)$ are the well known parabolic cylinder (Weber) functions
(see, e.g., \cite{AS68}) . Using the relations of these functions to the
confluent hypergeometric (Kummer) functions $M(\alpha,\beta;z)$ with
$M(\alpha,\beta;0)=1$, we have
\begin{widetext}
\begin{equation}
\vartheta_{1}(a,b)=\frac{\sqrt{2}\Gamma((1+b)/2)M((1+b)/2,1/2;a^{2}
/2)-2a\Gamma(1+b/2)M(1+b/2,3/2;a^{2}/2)}{\Gamma(b/2)M(b/2,1/2;a^{2}
/2)-\sqrt{2}a\Gamma((1+b)/2)M((1+b)/2,3/2;a^{2}/2)}\,, \label{theta_1}
\end{equation}
\end{widetext}
and the recursion relation
\begin{equation}
\vartheta_{l}(a,b)=(b+l-2)\,\vartheta_{l-2}(a,b)-a\,\vartheta_{l-1}(a,b)\,
\label{theta-recurs}%
\end{equation}
with $\vartheta_{0}(a,b)=1$.
We note two other special relations for later use:%
\begin{subequations}
\begin{align}
\vartheta_{l}(0,b)  &  =2^{l/2}\,\frac{\Gamma((l+b)/2)}{\Gamma(b/2)}%
\,,\label{theta-a=0}\\
\vartheta_{1}(a,b)  &  =b\,\sqrt{\pi/2}\,\,\mathrm{e}^{a^{2}/2}%
\operatorname{erfc}(a/\sqrt{2})+O(b^{2})\,. \label{theta-b=0}%
\end{align}
Here, $\operatorname{erfc}(x)$ denotes the complementary error function.

Now we are finally in the position to write down a scaling form for the
moments of the homogeneous density with known scaling functions. Collecting,
we obtain
\end{subequations}
\begin{equation}
\langle\Phi^{N}\rangle=(\hat{k}L^{d}/g^{2})^{-N/2}\,\vartheta_{N}(a,b) \, ,
\label{Momente}%
\end{equation}
with universal scaling functions $F_{N,0}(\{0\},a,b)=\vartheta_{N}(a,b)$ given to one-loop order
by Eq.~(\ref{theta_1}), or respectively, immediately following from
Eq.~(\ref{theta_1}) via the recursion relation~(\ref{theta-recurs}).

\section{Simulation results above $\bm{d_{\text{c}}}$}
\label{sim}

To complement our analytical calculations for $d > d_{\text{c}}$, we have
performed Monte Carlo simulations at $d=5$ of two critical models belonging to
the DP universality class (see~\cite{Hi00} and references therein). Naturally, the observables that we found best suited for our numerical work were not necessarily identical to those that are most convenient for doing field theory. In the following, we identify observables (ratios) that are convenient for numerical work, Eq.~(\ref{Ratios}). Then, we connect these observables with our field theoretic results which provides us with scaling functions for these observables, Eq.~(\ref{RatSkal}). We introduce a ratio $U$, Eq.~(\ref{Ratio_U}), which in a certain sense takes on the role in critical dynamics that the famous Binder cumulant plays in equilibrium critical phenomena. Equation~(\ref{U(a,b)}) gives our general analytical result for $U$. We derive the scaling form of the ubiquitous parameter $a$ at the critical point, Eq.~(\ref{afin}). This finally leads us to Eq.~(\ref{U0res}) for $U$, which is particularly well suited for comparison between theory and simulation.

We have simulated the contact process (CP) on simple cubic lattices
of size $L=4,8,16$ at the critical value of the respective control parameter
$\lambda$, $\lambda= \lambda_{\mathrm{c}}=1.13846(11)$, as well as the
site-directed percolation process (sDP) implemented via a generalized
Domany-Kinzel automaton~\cite{DoKi84,LuWi04} on bcc lattices of linear size
$L=8,16,32$ at the critical value of the occupation probability $p$, $p
=p_{\mathrm{c}}=0.0359725(2)$~\cite{Gra04}. In contrast to conventional
equilibrium simulation techniques, steady state finite-size quantities are
inaccessible for absorbing phase transitions at zero field because, close to
the transition point, the systems will be soon trapped in the absorbing state
without chance of escape. To circumvent these difficulties, we perform
simulations in non-zero source at criticality, as recently advocated
in~\cite{LuHe03}. In remainder of this section we will present the results of
our simulations and compare them to the analytic results derived in
Sec.~\ref{Scaling}. 

\begin{figure}[t]
\includegraphics[width=8.4cm,angle=0]{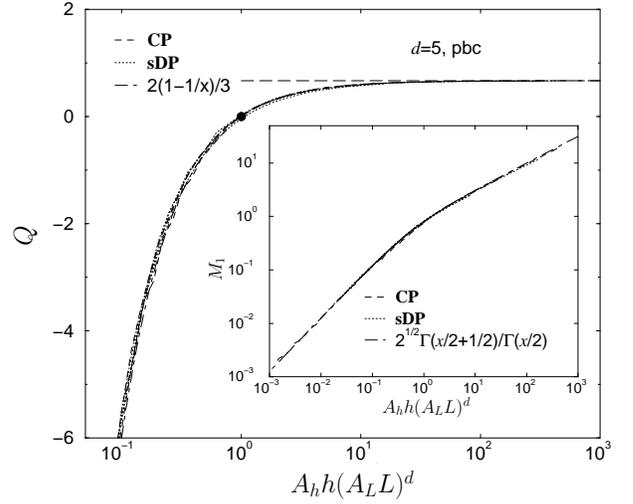}
\caption{ The universal order parameter scaling function $M_{1}$ (inset) and
the universal fourth order ratio scaling function $Q$ as a function of the
rescaled source.}%
\label{data}%
\end{figure}

Using first the lowest mode approximation without the $1$-loop corrections of
the higher modes we are interested in the moments of the order parameter, the
homogeneous density $\Phi$, about the absorbing state $\Phi=0$. According to
Eq.~(\ref{low-mod-scal}), we have the scaling equations
\begin{equation}
(A_{L}L)^{Nd/2}\langle\Phi^{N}\rangle=M_{N}\bigl(A_{\tau}\tau(A_{L}%
L)^{d/2},A_{h}h(A_{L}L)^{d}\bigr)\, . \label{MomSkal}%
\end{equation}
Deviating from the conventions used in Sec.~\ref{Scaling}, we have here
explicitly pulled the non-universal amplitudes $A_{L}$, $A_{h}$, and $A_{\tau
}$ out of the parameters $L$, $h$ and $\tau$, respectively.  In accord with our analytical
result~(\ref{theta-a=0}), we choose the normalizations $M_{1}(0,1)=\sqrt
{2/\pi}$ and $M_{4}(0,1)=3\bigl(M_{2}(0,1)\bigr)^{2}$. With these
normalizations, we get for the universal finite size scaling functions defined
in Eq.~(\ref{MomSkal}):
\begin{equation}
M_{N}(0,y)=2^{N/2}\,\frac{\Gamma\bigl((y+N)/2\bigr)}{\Gamma\bigl(y/2\bigr)}\,,
\label{M_N}%
\end{equation}
where bulk criticality, $\tau=0$, is assumed. For the order parameter, in
particular, this leads to the modified FSS scaling form
\begin{subequations}
\begin{align}
\langle\Phi\rangle=(A_{L}L)^{-d/2}M_{1}(0,A_{h}h(A_{L}L)^{d})\,,
\label{OP}%
\end{align}
with the universal scaling function
\begin{align}
M_{1}(0,y)  &  =\sqrt{2}\,\frac{\Gamma\bigl((y+1)/2\bigr)}{\Gamma
\bigl(y/2\bigr)}=\left\{
\begin{array}
[c]{ll}%
\sqrt{y}, & \qquad y\rightarrow\infty\\
\sqrt{{\pi}/{2}}\,\,y, & \qquad y\rightarrow0,
\end{array}
\right.  \,. \label{OP-Skal}%
\end{align}
For analyzing the numerical data, it is useful to define the ratios%
\end{subequations}
\begin{equation}
V=\frac{\langle\Phi^{2}\rangle}{\langle\Phi\rangle^{2}}-1\,,\quad
S=1-\frac{\langle\Phi^{3}\rangle}{3\langle\Phi\rangle\langle\Phi^{2}\rangle
}\,,\quad Q=1-\frac{\langle\Phi^{4}\rangle}{3\langle\Phi^{2}\rangle^{2}}\,.
\label{Ratios}%
\end{equation}
Note that the ratio $Q$ is identical in form to the well known Binder cumulant for equilibrium systems.
From Eqs.~(\ref{MomSkal}, \ref{M_N}), we immediately obtain
\begin{subequations}
\label{RatSkal}%
\begin{align}
V  &  =\frac{y\Gamma\bigl(y/2\bigr)^{2}}{2\Gamma\bigl((y+1)/2\bigr)^{2}%
}-1=\left\{
\begin{array}
[c]{ll}%
1/2y, & \qquad y\rightarrow\infty\\
y, & \qquad y\rightarrow0,
\end{array}
\right.  \,,\label{RatioV}\\[0.3cm]
S  &  =\frac{2}{3}\Big(1-\frac{1}{2y}\Big)\,,\qquad Q=\frac{2}{3}%
\Big(1-\frac{1}{y}\Big)\,, \label{RatioSQ}%
\end{align}
with the scaling argument $y=A_{h}h(A_{L}L)^{d}$.

Figure~\ref{data} compares our analytic results for the normalized order parameter $M_1$ and
the ratio ${Q}$ to our numerical findings. The solid dot marks the condition
$Q=0$ for $y=1$, and the horizontal dashed line corresponds to the limit
$2/3$. Fig.~\ref{data} demonstrates that the data of the lattice models obey
the modified FSS form Eqs.~(\ref{OP-Skal}) and (\ref{RatioSQ}), and that the
obtained scaling curves are in perfect agreement with the results of the
continuum theory. We rate this as an impressive manifestation of the
robustness of the DP universality class. Two further points are worth
stressing: (i) The order parameter assumes both asymptotic regimes
($M_{1}\simeq\sqrt{y}$ for $y\rightarrow\infty$ and $M_{1}\simeq\sqrt{\pi
/2}\,\,y$ for $y\rightarrow0$) predicted by our theory. (ii) As mentioned
above, the simulated systems got stuck quickly in the absorbing state if the
external source was turned off, $h=0$. Thus both, the analytical results as
well as the numerical simulations reflect that well-defined steady-states
exist close to the critical point for $h>0$ only.
\end{subequations}

The corrections due to the higher modes ($v>0$) and the exponential instead of
the algebraic crossover to mean field scaling for $y\rightarrow\infty$ are not
resolved by the numerical data. Note, however, that the leading terms of the
order parameter and second moment, cf.~Eqs.~(\ref{OP-Skal}) and (\ref{RatioV})
are correct in this limit. In this mean field region the order parameter
fluctuations are dominated by small Gaussian correlations. Hence, we have
$\langle\Phi^{N}\rangle/\langle\Phi\rangle^{N}=1+(N(N-1)/2)V+O(V^{2})$. Using
this expansion one easily demonstrates that $S$ and $Q$ as given in
Eq.~(\ref{RatioSQ}) show the correct asymptotic scaling including the
corrections $\propto y^{-1}$.

\begin{figure}[t]
\includegraphics[width=8.4cm,angle=0]{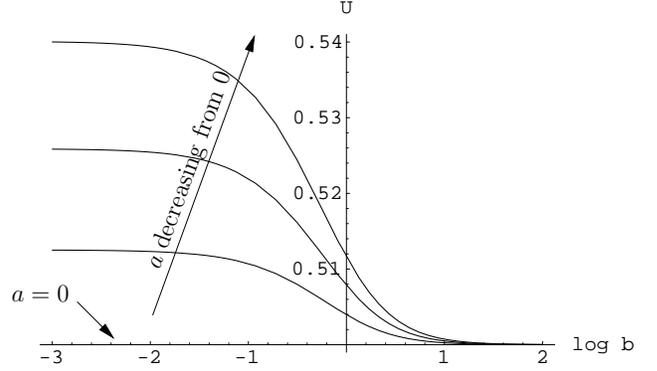}
\caption{The ratio $U$ as a function of $\ln b$ with $a$ as a parameter. For
$a=0$, $U$ lies on the abscissa, $U\equiv1/2$. For $a$ decreasing form zero,
the values of $U$ deviate increasingly from $1/2$ for $b\ll1$.}%
\label{Uab}%
\end{figure}

In the absorbing state, the ratios Eq.~(\ref{RatSkal}) are not finite for
$y\rightarrow0$. To analyze the scaling behavior in this limit, we introduce
the following combination of moments:
\label{Ratio_U_Funkt}%
\begin{equation}
U=\frac{\langle\Phi^{2}\rangle\langle\Phi^{3}\rangle-\langle\Phi\rangle
\langle\Phi^{2}\rangle^{2}}{\langle\Phi\rangle\langle\Phi^{4}\rangle
-\langle\Phi\rangle\langle\Phi^{2}\rangle^{2}}=\frac{2-3S}{2-3Q}\,,
\label{Ratio_U}%
\end{equation}
which can be viewed as an analog in critical dynamics of the famous Binder cumulant.
Inserting the lowest-mode scaling functions, Eq.\thinspace(\ref{RatioSQ}),
this ratio becomes simply a constant equal to $1/2$. This value is indeed
correct in the limit $y\rightarrow\infty$, but for $y\rightarrow0$, we should
expect deviations due to the finite-size shift of the critical control
parameter $\tau$. Using the scaling form~(\ref{Momente}) for the order
parameter moments, we obtain%
\begin{equation}
U(a,b)=\frac{\vartheta_{2}(a,b)\vartheta_{3}(a,b)-\vartheta_{1}(a,b)\vartheta
_{2}(a,b)^{2}}{\vartheta_{1}(a,b)\vartheta_{4}(a,b)-\vartheta_{1}%
(a,b)\vartheta_{2}(a,b)^{2}}\,, \label{U(a,b)}%
\end{equation}
as functions of the parameters $a$ and $b$, with $\vartheta_{l}(a,b)$
following from Eqs.~(\ref{theta_1}) and (\ref{theta-b=0}) via the recursion
relation~(\ref{theta-recurs}). $U(a,b)$ is displayed in Fig.~\ref{Uab} as a
function of $\ln b$ with $a$ as a parameter. It is only in the case $a=0$ that
$U$ is constantly equal to $1/2$, whereas $U$ deviates and grows from this
value for $b\ll1$ when $a$ gets increasingly negative. Thus, the limit
$U_{0}(a)=\lim_{b\rightarrow0}U(a,b)$ at the bulk critical point is a
convenient measure of the shift of the critical control parameter due to
finite size.

Now, let us look more closely at the parameter $a$ as given in Eq.~(\ref{a}).
For $x=y=0$ we get%
\begin{equation}
a=-\Big[1-\frac{7v^{2}}{4}D^{(2)}(0)+O(v^{4})\Big]\,\Big\{1+O(v^{2}%
)\Big\}vD^{(1)}(0)\,. \label{a-krit}%
\end{equation}
Defining a new nonuniversal length $L_{0}$ by the relation%
\begin{equation}
vD^{(1)}(0)=(L_{0}/L)^{d/2-2}\,, \label{L-null}%
\end{equation}
we obtain%
\begin{align}
a=&-\Big[1-\frac{7D^{(2)}(0)}{4D^{(1)}(0)^{2}}(L_{0}/L)^{d-4}+O((L_{0}%
/L)^{2d-8})\Big]
\nonumber \\
&\times \Big\{1+O((L_{0}/L)^{d-4})\Big\}(L_{0}/L)^{d/2-2}\,.
\label{avon L}%
\end{align}
At first glance, the correction factor $-7D^{(2)}(0)/4D^{(1)}(0)^{2}=-2.096$
in $d=5$ seems to be a universal contribution. Note, however, that this
correction factor merely represents the 1-loop contribution and that the
2-loop contribution $O((L_{0}/L)^{d-4})$ of the second factor in
Eq.~(\ref{avon L}) is of the same order in $L_{0}/L$ as the $1$-loop
contribution. Therefore, to be consistent, one either has to take only the
lowest order in Eq.~(\ref{avon L}), or, if one seeks to proceed to next to leading order, one has to account for the 2-loop contribution to the finite-size shift of the control parameter $\tau$.
This subtlety was overlooked by Chen and Dohm \cite{CheDo98} in their work on
FFS in the Ising model, and their derivation of universal scaling functions.  Because there exists to date
no 2-loop calculation of the shift of the critical control parameter, which
would eventually lead to a universal correction proportional to $(L_{0}%
/L)^{d-4}$, we introduce here a wild-card $K$ for this universal correction.
The introduction of $K$ produces
\begin{equation}
a=-\Big[1+K\,(L_{0}/L)^{d-4}+O((L_{0}/L)^{2d-8})\Big](L_{0}/L)^{d/2-2}\,,
\label{afin}%
\end{equation}
with $L_{0}$ and $K$ to be determined by fits to the numerical data.

\begin{figure}[t]
\includegraphics[width=8.4cm,angle=0]{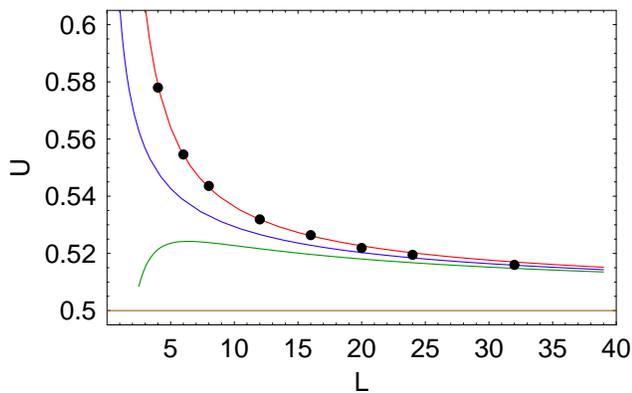}
\caption{ (Color online) The ratio $U$ as a function of $L$. The meaning of the various curves is explained in the text.}
\label{fig:ratU}
\end{figure}

Next, we revisit the ratio $U$. From the representations~(\ref{theta-recurs})
(\ref{theta-b=0}) of the functions $\vartheta_{l}$ in Eq.~(\ref{U(a,b)}), we
obtain in the limit $b\rightarrow0$
\begin{subequations}
\label{U0res}
\begin{align}
U &= U_{0}(a): =U(a,0)\nonumber\\
&=\frac{\bigl[F(a)-a\bigr]\bigl[(1+a^{2})-aF(a)\bigr]}
{(2+a^{2})F(a)-a(3+a^{2})}\,, \label{Ratio_U_0}
\end{align}
where 
\begin{equation}
F(a)=\sqrt{2/\pi}\exp(-a^{2}/2)\operatorname{erfc}(a/\sqrt{2})^{-1},
\label{F_a}%
\end{equation}
\end{subequations}
with ${a=\sqrt{L_{0}/L}\,\bigl(1+KL_{0}/L+O((L_{0}/L)^{2})\bigr)}$. The ratio
$U$ is shown for $d=5$ in Fig.~\ref{fig:ratU}. The solid dots stem from our Monte-Carlo simulations of critical sDP on bcc lattices of linear size $L=4$ to $L=32$. The red upper curve is a fit to the numerical data with $L_{0}=1.01$ and $K=2.17$. As expected, the nonuniversal length scale $L_{0}$ is of the order of the lattice spacing. Note that the numerical result for correction parameter $K$ is positive whereas the pure 1-loop calculation, which entirely misses the $O((L_{0}/L)^{d-4})$-term in Eq.~(\ref{avon L}), pretends a negative value of $-2.096$. For the purpose of demonstration, we include in Fig.~\ref{fig:ratU} the curve of $U_0$ pertaining to this flawed value of $K$, green lower curve. Note that this curve has a
non-physical maximum near $L=6$, which clearly shows that a pure 1-loop calculation is incomplete and misleading and which underscores our previous reasoning that
$2$-loop contributions to the shift of the control parameter cannot be
neglected for the interpretation of the numerical data. For further
comparison, we also plot $U_0$ using the correct 1-loop
result for the control parameter shift, i.e, with $K$ set to zero (blue middle
curve). The figure shows that the corrections cannot be neglected below
$L\thickapprox24$ due to the slow decrease of $a\sim\sqrt{L_{0}/L}$ with
increasing $L$. The zero-mode limit $1/2$ (brown straight
line) approximates $U$ reliably for only very large values of $L$, which were out of reach for our simulations.

\section{Crossover to mean field behavior}
\label{crossoverToMeanField} \noindent

In this section, we consider the crossover to the mean field behavior in
the infinite volume limit $w\rightarrow\infty$ in spatial dimensions $d>4$. 
If $w$ is comparable with or greater than $1$, i.e., for $r=(\tau-\tau_{c}%
)+M\geq(2\pi/L)^{2}$, we can and do calculate the order parameter
$M=\langle\Phi\rangle=g\langle s(\mathbf{r},t)\rangle$ and its correlation
$\chi^{-1}=\langle\delta\Phi^{2}\rangle=\langle\Phi^{2}\rangle-\langle
\Phi\rangle^{2}=g^{2}L^{-d}\int d^{d}r\,\langle s(\mathbf{r},t)s(\mathbf{0}%
,t)\rangle^{\mathrm{cum}}$ in a standard $1$-loop perturbation expansion based
on functional integrals with weight $\exp(-\mathcal{J})$, where $\mathcal{J}$
is the response functional as given in
Eq.~(\ref{eq:action_reggeon_field_theory}). The results for
$M$ and $\chi^{-1}$ produced by this direct calculation are then compared with the corresponding expressions
calculated with the steady state distribution function, Eq.~(\ref{Stat_Vert}).
This comparison reveals that neither the lowest-mode approximation nor the
$1$-loop calculation using the Markovian approximation capture the correct
crossover behavior as produced by the direct calculation. Equation~(\ref{change}) nails down the difference in the outcome of the direct calculation and the one that uses the Markovian approximation.

For calculating $M$ and $\chi^{-1}$ without recourse to the dynamic free
energy, we need to know both the propagator and the correlator implied in
$\mathcal{J}$. Applying the shift $s\rightarrow s+M/g$ to $\mathcal{J}$, we
readily obtain
\begin{subequations}
\begin{align}
G(\mathbf{q},t)  &  =\theta(t)\exp\bigl(-\lambda(r+\mathbf{q}^{2}%
)t\bigr)\,,\label{G}\\
C(\mathbf{q},t)  &  =\frac{M}{2(r+\mathbf{q}^{2})}G(\mathbf{q},\left\vert
t\right\vert )\, , \label{C}%
\end{align}
as the propagator and correlator in time and momentum space. Then, to $1$-loop
order, the equation of state follows as
\end{subequations}
\begin{equation}
H=\tau M+\frac{1}{2}M^{2}+\frac{g^{2}}{2}L^{-d}\sum_{\mathbf{q}}%
C(\mathbf{q},0)\,. \label{H}%
\end{equation}
To the same order we obtain for the correlation
\begin{widetext}
\begin{align}
\frac{L^{d}}{g^{2}}   \chi^{-1}&=C(\mathbf{0},0)  +(\lambda g)^{2}L^{-d}\sum_{\mathbf{q}}\iint\limits_{-\infty}^{\quad
\;\;0}dt\,dt^{\prime}\,\Big[G(\mathbf{0},-t)G(\mathbf{0},-t^{\prime
})\Big(\frac{1}{2}C(\mathbf{q},t-t^{\prime})^{2}-2C(\mathbf{q},t-t^{\prime
})G(\mathbf{q},t-t^{\prime})\Big)\nonumber\\
&  \qquad\qquad\qquad\qquad\qquad\qquad\qquad+G(\mathbf{0},-t)C(\mathbf{0},-t^{\prime}%
)\Big(2C(\mathbf{q},t-t^{\prime})G(\mathbf{q},t-t^{\prime})-G(\mathbf{q}%
,t-t^{\prime})^{2}\Big)\Big]\,. \label{chi}%
\end{align}
\end{widetext}
After some calculation and after rescaling of the parameters and fields as
before, we get the equation of state,
\begin{equation}
y=(x+z)\Big(1+\frac{1-wD^{(1)}(w)}{(x+z)^{2}}\Big)\frac{z}{2}-\frac{z^{2}}
{4}\,, \label{ZustGl}
\end{equation}
in terms of the scaling variables defined by Eqs.~(\ref{x_and_y}) and
(\ref{w_und_m}). The correlations in terms of these variables are given by
\begin{align}
\frac{2L^{d}}{g^{2}}\chi^{-1}= &\! \frac{z}{x+z}\bigg(1-\frac{2-4wD^{(1)}
(w)+3wD^{(1)}(3w/2)}{(x+z)^{2}}
\nonumber \\
&\qquad\qquad\qquad +z\frac{1-w^{2}D^{(2)}(w)}{(x+z)^{3}}\bigg)\,.
\label{Korrel}%
\end{align}
If $w\gg1$, we have $D^{(l)}(w)\simeq w^{-l}$ up to exponentially small
corrections. Using these properties, we find that the equation of state and
the correlation approach their mean field forms in the infinite volume limit
with exponentially decreasing deviations. In contrast to this exponential
crossover, the lowest-mode approximation, which corresponds to letting
$w^{l}D^{(l)}(w)\rightarrow0$, produces unphysical algebraic
crossover to mean field behavior with decreasing deviations proportional to
$(x+z)^{-2}$.

Recall that we have calculated in Sec.~\ref{Scaling} with the help of the
steady state distribution a scaling form for the moments of $\Phi$,
Eq.~(\ref{Momente}). This equation implies scaling forms for the equation of
state and the correlations, which we in the following wish to compare to
Eqs.~(\ref{ZustGl}) and (\ref{Korrel}). For simplicity, we focus on the
following three regions of phase space: the absorbing phase region $x\gg1$,
the active phase region $-x\gg1$, both with small source $x^{2}\gg4y$, and the
region with large source including the bulk critical point $4y\gg x^{2}$.
Expanding Eqs.~(\ref{ZustGl}) and (\ref{Korrel}) for $x\gg1$, $x^{2}\gg4y$, we
obtain
\begin{subequations}
\label{ober}%
\begin{align}
\frac{L^{d/2}}{g}M &\simeq\frac{y}{x}\bigg(1-\frac{y}{x^{2}}-\frac
{1-wD^{(1)}(w)}{x^{2}}\bigg)\,,\label{M_ob}\\
\frac{L^{d}}{g^{2}}\chi^{-1} &\simeq\frac{y}{x^{2}}\bigg(1-\frac{3y}{x^{2}}\nonumber\\
&-\frac{3-5wD^{(1)}(w)+3wD^{(1)}(3w/2)}{x^{2}}\bigg). \label{chi_ob}
\end{align}
In the active region $-x\gg1$, $x^{2}\gg4y$, we get
\end{subequations}
\begin{subequations}
\label{unter}%
\begin{align}
&\frac{L^{d/2}}{g}M   \simeq\left\vert x\right\vert \bigg(1+\frac{y}{x^{2}%
}-\frac{1-wD^{(1)}(w)}{x^{2}}\bigg)\,,\label{M_unt}\\
&\frac{L^{d}}{g^{2}}\chi^{-1}    \simeq1-\frac{y}{x^{2}}
\nonumber \\
&+\frac{1+3wD^{(1)}
(w)-3wD^{(1)}(3w/2)-2w^{2}D^{(2)}(w)}{x^{2}}\,. \label{chi_unt}%
\end{align}
Finally, we find for $4y\gg x^{2}$
\end{subequations}
\begin{subequations}
\label{mitte}%
\begin{align}
&\frac{L^{d/2}}{g}M   \simeq\sqrt{y}\bigg(1-\frac{x}{2\sqrt{y}}-\frac
{1-wD^{(1)}(w)}{4y}\bigg)\,,\label{M_mit}\\
&\frac{L^{d}}{g^{2}}\chi^{-1}    \simeq\frac{1}{2}-\frac{x}{4\sqrt{y}}
\nonumber\\
&-\frac{1-4wD^{(1)}(w)+3wD^{(1)}(3w/2)+w^{2}D^{(2)}(w)}{8y}\,. \label{chi_mit}%
\end{align}
Next, let us see what our steady state distribution, Eq.~(\ref{Stat_Vert}),
tells us, and let us compare that to the above. Using the asymptotic
properties of the parabolic cylinder functions~\cite{AS68} in the three
regions, we obtain for $M$ the same expressions as displayed in Eqs.~(\ref{M_ob}),
(\ref{M_unt}) and (\ref{M_mit}). For the correlations, we recover Eqs.~(\ref{chi_ob}),
(\ref{chi_unt}) and (\ref{chi_mit}) up to one alteration: the
function $D^{(1)}(3w/2)$ is replaced by%
\end{subequations}
\begin{equation}
D^{(1)}(3w/2)\rightarrow D^{(1)}(w)-\frac{w}{2}D^{(2)}(w)\,, \label{change}%
\end{equation}
which is an identity to linear order in $w$ but which modifies the correlations at higher orders.
The mean-field parts of the expressions for $\chi^{-1}$, given by the
respective first two terms on the right hand sides of Eqs.~(\ref{chi_ob}),
(\ref{chi_unt}) and (\ref{chi_mit}), are identical in both approaches. For $w \gg1$, where we have $D^{(l)}(w)\simeq w^{-l}$ up to exponentially small corrections, Eqs.~(\ref{ober}) to (\ref{mitte}) tend to the mean-field behavior with exponentially decaying corrections. After the
replacement~(\ref{change}) (i.e., in the approach based on the steady state
distribution), however, these corrections for the correlation
$\chi^{-1}$ fall off only algebraically. This incorrect feature
is a consequence of the Markovian approximation as the direct
calculation shows.

\section{Finite size effects below $\mathbf{d_{\text{c}}}$}
\label{belowDc}

As mentioned above, in a former publication \cite{JSS88}, one of us and
coworkers have calculated finite size effects for absorbing nonequilibrium
processes belonging the DP universality class in spatial dimensions
$d=4-\varepsilon<4$. There, systems without a source were considered, and the
consequences of the finite size scaling for the relaxation behavior were
scrutinized. Here, we are interested in the steady state properties in the
presence of the source $h$. We calculate various quantities in an $\varepsilon$-expansion, most notably the parameters $a$ and $b$ and the ratio $U$. Equation~(\ref{a_und_b_unt}) gives our $\varepsilon$-expansion results for $a$ and $b$, and Eq.~(\ref{U_eps}) states our $\varepsilon$-expansion result for $U$.

A $1$-loop calculation for $d<4$ can be done in much the same way as the
calculation for $d>4$ presented in Sec.~\ref{Dyn Freie Energie}. Now, however,
in addition to the functions $D^{(l)}(w)$ in Eqs.~(\ref{Par2}), the brackets
in Eqs.~(\ref{Par1}) become IR divergent, and therefore, they no longer can be
simply included in non-universal amplitudes. Rather, these additional
divergencies must be handled with a renormalization procedure and the
renormalization group equation. For general background on these methods, we
refer to the usual textbooks, e.g.\ \cite{ZJ96,Am84}; for applications of
these techniques to the DP universality class consult, e.g.,
Refs.~\cite{Ja81,Ja01,JaTa05}.

Whereas the coupling constant $g$ is dangerously irrelevant in $d > 4$, it is, respectively, marginal and relevant in $d=4$ and $d<4$. Therefore, it is useful for the case $d<4$ presented in this section and the case $d=4$ to be presented in Sec.~\ref{atDc} to recast the dynamic free energy~(\ref{vollst_freie_energie}) as
\begin{equation}
\Sigma[\tilde{S}, S]=\lambda L^{d}\int\mathrm{d}%
t\,\Big\{\tilde{S}\Big[\lambda^{-1}\hat{k}\partial_{t}+\hat{\tau}%
+\frac{\hat{g}}{2}(S-{\tilde{S})}\Big]S-h{\tilde{S}}\Big\},
\label{vollst_freie_energie_recast}%
\end{equation}
with $S(t)=L^{-d}\int\mathrm{d}^{d}r\,s(\mathbf{r},t)$ and likewise for $\tilde{s}$.
To facilitate the renormalization procedure and to cleanly keep track of bare
(unrenormalized) and renormalized quantities, we henceforth label bare fields
and parameters with a ring $\mathring{}\,$, i.e., we let $s\rightarrow
\mathring{s}$, $\tilde{s}\rightarrow\mathring{\tilde{s}}$, $\tau
\rightarrow\mathring{\tau}$, and so on, and we reserve symbols without a ring
$\mathring{}\,$ for their renormalized counterparts. The bare and the
renormalized quantities are related via the renormalization scheme
\begin{subequations}
\label{Renorm}%
\begin{align}
\mathring{s}  &  =Z^{1/2}s\,,\qquad\mathring{\tilde{s}}=Z^{1/2}\tilde
{s}\,,\qquad\mathring{h}=Z_{\lambda}^{-1}Z^{1/2}h\,,\\
\mathring{\lambda}  &  =Z_{\lambda}Z^{-1}\lambda\,,\qquad\mathring{g}%
^{2}=Z_{\lambda}^{-2}Z^{-1}Z_{u}g^{2}\,,\\
\mathring{\tau}  &  =Z_{\lambda}^{-1}Z_{\tau}\tau+\mathring{\tau}_{c}\,.
\end{align}
The renormalization factors $Z$, $Z_{\tau}$ and so on are determined as to
eliminate the $\varepsilon$-poles arising in a dimensional regularized
calculation of the momentum space integrals. This kind of calculation orders naturally in powers of
a dimensionless coupling constant $u$ defined by $u=G_{\varepsilon}%
\mu^{-\varepsilon}g^{2}$, where $\mu^{-1}$ is a convenient length scale, and
$G_{\varepsilon}=\Gamma(1+\varepsilon/2)/(4\pi)^{d/2}$. The renormalization factors are given to $1$-loop by%
\end{subequations}
\begin{subequations}
\label{oneLoopFactors}
\begin{align}
Z  &  =1+\frac{u}{4\varepsilon}\,,\qquad Z_{\lambda}=1+\frac{u}{8\varepsilon
}\,,\\
Z_{\tau}  &  =1+\frac{u}{2\varepsilon}\,,\qquad Z_{u}=1+\frac{2u}{\varepsilon
}\,.
\end{align}
With help of the renormalization scheme~(\ref{Renorm}) and the renormalization factors~(\ref{oneLoopFactors}), we find that the renormalized versions of the parameter functions featured in the dynamic free energy~(\ref{vollst_freie_energie_recast}) are given by
\end{subequations}
\begin{subequations}
\label{Par3}%
\begin{align}
\hat{k}  = &\, \Big[1-\frac{u}{4}\ln\Big(\frac{\mu L}{2\pi}\Big)+\frac{u}%
{8}\sigma^{\prime}(w)\Big]\,,\\
\hat{\tau}  = &\, \Big[1-\frac{u}{2}\ln\Big(\frac{\mu L}{2\pi}\Big)+\frac{u}
{4}\sigma^{\prime}(w)\Big]\tau
\nonumber\\
&\qquad\quad +\frac{u}{4}\Big[\sigma(w)-w\sigma^{\prime
}(w)\Big]\Big(\frac{2\pi}{L}\Big)^{2}\,,\\
\hat{g}  = &\, \Big[1-u\ln\Big(\frac{\mu L}{2\pi}\Big)+\frac{u}{2}\sigma
^{\prime}(w)\Big]g \,.
\end{align}
\end{subequations}
Here we have defined the function%
\begin{equation}
\sigma(w)=w(\ln w-1)-\frac{1}{\pi^{2}}D^{(1)}(w)\,,
\end{equation}
where it is understood that $D^{(1)}(w)$ is taken at $d=4$ and where
$\sigma^{\prime}(w)$ stands, as usual, for the derivative of $\sigma(w)$. The virtue of the
function $\sigma(w)$ is that it and its derivative lack the non-analytic
behavior of $D^{(l)}(w)$ for $w\rightarrow0$. However, as shown in the
appendix, these functions are nevertheless logarithmically divergent in the
bulk limit $w\rightarrow\infty$. In principle, one should handle these
divergences by subtracting a term $w\ln(1+w)$ or $\ln(1+w)$, respectively, as
done in \cite{JSS88}. These subtractions, with $w$ as given by
Eq.~(\ref{Def_w}), combine with the logarithm in the first brackets of
Eqs.~(\ref{Par3}) to produce the IR-divergent term $\ln\bigl((2\pi/\mu
L)^{2}+(\tau+M)/\mu^{2}\bigr)$, which should be eliminated by the
renormalization flow. Nonetheless, we can here set these subtleties aside and
ignore the divergences for $w\rightarrow\infty$ because we are only interested
in the strong finite size case $w\ll1$.

The perturbation results for the parameter functions, Eqs.~(\ref{Par3}), cannot be used directly as they stand. These results must be transported by the renormalization group flow to a non-critical
region. To this end, we derive in a standard fashion Gell-Mann--Low renormalization group equations (RGEs) for the parameter functions via exploiting the fact that the bare theory must be independent of the length scale $\mu^{-1}$ introduced by renormalization:
\begin{subequations}
\label{RGEs}%
\begin{align}
\mathcal{D}_\mu \ln \hat{k}  &  =\gamma \,,
\\
\mathcal{D}_\mu \ln \hat{\tau}  &  = \gamma  - \zeta
\,,\\
\mathcal{D}_\mu \ln \hat{g}  &  = \textstyle{\frac{1}{2}} (3\gamma - 2 \zeta) \,,
\end{align}
\end{subequations}
where, $\mathcal{D}_\mu$ stands for the renormalization group differential operator
\begin{align}
\mathcal{D}_\mu = \mu \partial_\mu + \beta \partial_u +  \lambda \zeta \partial_\lambda + \tau \kappa  \partial_\tau + \frac{M}{2} \left( \varepsilon - \frac{\beta}{u} - \gamma \right) \partial_M \,,
\end{align}
and where $\gamma$, $\zeta$ and so on are the usual RG functions. For DP, these RG functions are known to 2-loop order~\cite{Ja81,Ja01}:
\begin{subequations}
\label{RGFunkt}%
\begin{align}
\gamma &  =-\frac{u}{4}+\Big(2-3\ln\frac{4}{3}\Big)\frac{3u^{2}}{32}%
\,,\\
\zeta &  =-\frac{u}{8}+\,\Big(17-2\ln\frac{4}{3}\Big)\frac{u^{2}}{256}%
 \,,\\
\kappa &  =\frac{3u}{8}-\,\Big(7+10\ln\frac{4}{3}\Big)\frac{7u^{2}}%
{256} \,,\\
\beta &  =-\varepsilon u+\frac{3u^{2}}{2}-\Big(169+106\ln\frac{4}{3}%
\Big)\frac{u^{3}}{128} \,.
\end{align}
\end{subequations}
where we have included the 2-loop contributions, even though we work in this section only to 1-loop order, because we will need them in Sec.~\ref{atDc}. The RGEs can be solved using the method of characteristics. The idea behind this method is to consider all the scaling parameters as a function of a single flow parameter $\ell$. One sets up characteristic equations that describe how the scaling parameters transform under a change of $\ell$. The characteristic for the inverse length scale $\mu$ is trivial and has the solution $\bar{\mu}(\ell)=\mu \ell$, i.e., a change of $\ell$ corresponds to a change of the external inverse length scale. With help of the solution to the remaining characteristics and also with help of a dimensional analysis to account for naive dimensions, we obtain
\begin{subequations}
\label{parameterFktsScalingFormsGen}
\begin{align}
\hat{k} (\tau, M, u, \mu, L) &  = X(\ell)^{-1} \hat{k}\!  \left(\frac{\bar{\tau}(\ell)}{(\mu \ell)^2}, \frac{\bar{M}(\ell)}{(\mu \ell)^2}, \bar{u} (\ell) , 1, \mu \ell L \right) ,
\\
\hat{\tau} (\tau, M, u, \mu, L) &  = (\mu \ell)^2 X(\ell)^{-1} X_{\lambda} (\ell)
\nonumber \\
&\! \times \hat{\tau}\!  \left(\frac{\bar{\tau}(\ell)}{(\mu \ell)^2}, \frac{\bar{M}(\ell)}{(\mu \ell)^2}, \bar{u} (\ell) , 1, \mu \ell L \right) ,
\\
\hat{g} (\tau, M, u, \mu, L) &  = (\mu \ell)^{\varepsilon/2}  X(\ell)^{-3/2} X_{\lambda} (\ell)
\nonumber \\
&\times \hat{g}\!  \left(\frac{\bar{\tau}(\ell)}{(\mu \ell)^2}, \frac{\bar{M}(\ell)}{(\mu \ell)^2}, \bar{u} (\ell) , 1, \mu \ell L \right) ,
\end{align}
\end{subequations}
where
\begin{subequations}
\label{parameterFktsScalingFormsIngredients1}
\begin{align}
\bar{\tau}(\ell) &= \tau X_{\tau} (\ell) \, ,
\\
\bar{M}(\ell) &= M \ell^{\varepsilon/2} [\bar{u} (\ell)/u]^{1/2} X(\ell)^{-1/2} .
\end{align}
\end{subequations}
At this stage, the scaling relations~(\ref{parameterFktsScalingFormsGen}) are still rather formal because we still must determine $X(\ell)$, $X_{\lambda} (\ell)$, $X_{\tau} (\ell)$ and $\bar{u}(\ell)$ by solving their respective characteristic. The characteristic for the dimensionless coupling constant $u$ is given by
\begin{equation}
\ell \frac{d\upsilon}{d\ell }=\beta(\upsilon)\, \label{Char-u}%
\end{equation}
where we abbreviated $\upsilon=\bar{u}(\ell )$. The remaining characteristics are
all of the same structure:
\begin{equation}
\label{charQ}
\ell \frac{d\ln Q(\upsilon)}{d\ell }=q(\upsilon)\,.
\end{equation}
Here, $Q$ is a placeholder for $X$, $X_{\tau}$, and $X_{\lambda}$,
respectively, and $q$ is a placeholder for $\gamma$, $\kappa$, and $\zeta$,
respectively. As usual, solving the characteristics leads to qualitatively different results depending on whether we consider the upper critical dimension or dimensions below it. We will return to $d=4$ in Sec.~\ref{atDc}.

For $d<4$, the dimensionless coupling constant $u$ flows to the stable fixed point $u_{\ast}=2\varepsilon
/3+O(\varepsilon^{2})$, and, consequently, $X(\ell)$ etc.\ display power law behavior described by the well known critical exponents of the DP universality class. Using a compact notation where $\hat{p}$ stands ambiguously for $\hat{k}$, $\hat{\tau}$ and $\hat{g}$, we can write the resulting scaling form for the parameter functions as
\begin{equation}
\hat{p}(\tau,M,L)=\ell^{\delta_{\hat{p}}}\hat{p}(\ell^{-1/\nu}\tau,\ell^{-\beta/\nu
}M,\ell L)\,,
\end{equation}
with $\delta_{\hat{k}}=-\eta=d-2\beta/\nu$, $\delta_{\hat{\tau}} = z-\eta=\gamma/\nu$ and $\delta_{\hat{g}} =(2z-d-3\eta )/2=(\gamma-\beta)/\nu$, respectively. Here, the three independent critical exponents are
given by
\begin{equation}
\beta=1-\frac{\varepsilon}{6}\,,\qquad\gamma=1+\frac{\varepsilon}{6}%
\,,\qquad\nu=\frac{1}{2}+\frac{\varepsilon}{16}\,, \label{Exponenten}%
\end{equation}
up to terms of order $\varepsilon^{2}$ \cite{Ja81,Ja01}. The exponents $\beta$ and $\gamma$ must not be confused with the RG functions discussed above. Now, we choose the flow parameter
$\ell=2\pi/\mu L\ll1$ to eliminate the IR-diverging logarithm $\ln\bigl(\mu
L/2\pi\bigr)$. The parameter $\ell$ must be small to reach the asymptotic
region, i.e., to produce universal behavior. Of course, this is a condition on
the size $L$, which must not be small in comparison to a non-universal length
scale $L_{0}$ which is set in our simulations by the lattice constant. After
implementing our choice of $\ell$, we obtain the basic parameter functions in
scaling form
\begin{subequations}
\label{Par4}%
\begin{align}
\hat{k}  = &\, \bigl(\mu L/2\pi\bigr)^{2\beta/\nu-d}\Big[1+\frac{\varepsilon
}{12}\sigma^{\prime}(w)\Big]\,,\\
\hat{\tau}  = &\, \bigl(\mu L/2\pi\bigr)^{-\gamma/\nu}\Big\{\Big[1+\frac
{\varepsilon}{6}\sigma^{\prime}(w)\Big]\tau\bigl(\mu L/2\pi\bigr)^{1/\nu}\nonumber\\
&\qquad\qquad\qquad +\frac{\varepsilon}{6}\Big[\sigma(w)-w\sigma^{\prime}(w)\Big]\mu^{2}\Big\}\,,\\
\hat{g}  = &\, \bigl(\mu L/2\pi\bigr)^{(\beta-\gamma)/\nu}\Big[1+\frac
{\varepsilon}{3}\sigma^{\prime}(w)\Big] g \,,
\end{align}
where now
\end{subequations}
\begin{equation}
w=\bigl[\tau\bigl(\mu L/2\pi\bigr)^{1/\nu}+M\bigl(\mu L/2\pi\bigr)^{\beta/\nu
}\bigr]/\mu^{2}\,.
\end{equation}

\begin{figure}[t]
\includegraphics[width=8.4cm,angle=0]{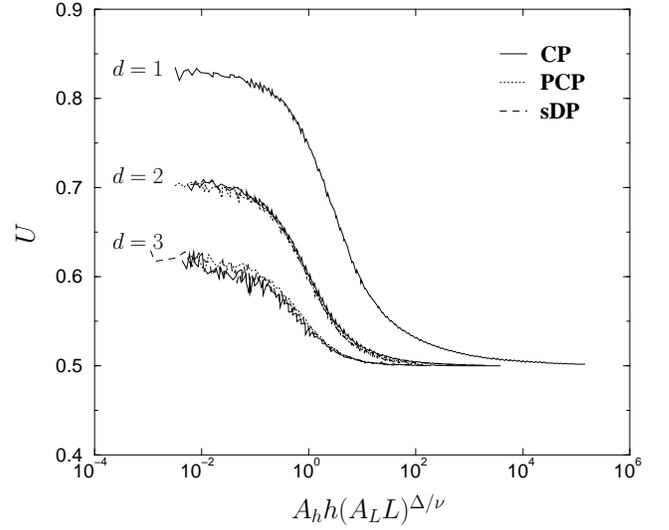}
\caption{The universal ratio $U$ at the bulk critical point as a function of
the scaled source $h/L^{{\Delta}/\nu}$ in dimensions $d=1,2,3$.}%
\label{U_von_d}%
\end{figure}

Next, let us return to the parameters $a$ and $b$ of the dynamic free
energy~(\ref{resc_free_energy}) and the steady state
distribution~(\ref{Stat_Vert}). Because we are interested in the strong finite
size case $w\ll1$, we can approximate $\sigma(w)\approx\sigma(0)+w\sigma
^{\prime}(0)$ and $\sigma^{\prime}(w)\approx\sigma^{\prime}(0)$, where $\sigma (0) = - 8 \ln 2 /\pi^2 \approx -0.56184$ and $\sigma^{\prime}(0) = -1 - C_E - 2\ln 2/3 -6\zeta^\prime(2)/\pi^2 \approx -1.85789$ with $C_E$ and $\zeta$ denoting Euler's constant and Riemann's $\zeta$-function, respectively.  Recalling the definitions of $a$ and $b$, Eqs.~(\ref{abDef}), we find after some algebra their $\varepsilon$-expansions to be given by
\begin{subequations}
\label{a_und_b_unt}%
\begin{align}
a  = &\, \pi\sqrt{6/\varepsilon}\,\bigl[1-A\varepsilon+O(\varepsilon
^{2})\bigr](\tau/\mu^{2})(\mu L/2\pi)^{1/\nu}
\nonumber\\
&\qquad\qquad -\,\frac{8\ln2}{\pi}%
\sqrt{\varepsilon/6}\bigl[1+O(\varepsilon)\bigr]\,,\\
b  = & \,(3\pi^{2}/\varepsilon)\,\bigl[1-2A\varepsilon+O(\varepsilon
^{2})\bigr](H/\mu^{4})(\mu L/2\pi)^{\Delta/\nu}\,,
\end{align}
where $A=(\ln\pi)/4-(\ln2)/12+(C_{E}-1)/8-3\zeta^{\prime}(2)/(2\pi)^{2}%
\approx0.24688$ and $\Delta=\beta+\gamma$. Now, we are finally in the position to address our main observable, the momenta ratio $U_{0}(a)=U(a,b\rightarrow0)$. Expanding
Eqs.~(\ref{U0res}) with $a$ as given in Eq.~(\ref{a_und_b_unt}) in $\varepsilon$ we find that
\end{subequations}
\begin{align}
U_{0}=\,\frac{1}{2}+\sqrt{\frac{\varepsilon}{3\pi}}\Big(\frac{4}{\pi}
-1\Big)\ln2+\frac{2\varepsilon}{\pi}\Big(&1-\frac{4}{3\pi}-\frac{16}{3\pi^{2}
}\Big)\bigl(\ln2\bigr)^{2}
\nonumber\\
&+O(\varepsilon^{3/2}) \label{U_eps}%
\end{align}
at the bulk critical point $\tau=0$. Setting $\varepsilon=1$, $2$, and $3$ in
Eq.~(\ref{U_eps}) we obtain estimates of $U_{0}$ for systems belonging to the
DP universality class in spatial dimensions $3$, $2$, and $1$, respectively:%
\begin{equation}
U_{0}=\left\{
\begin{array}
[c]{ccc}%
0.573\qquad & \text{for} & \qquad d=3\\
0.609\qquad & \text{for} & \qquad d=2\\
0.639\qquad & \text{for} & \qquad d=1
\end{array}
\right.  \,. \label{U_d}%
\end{equation}
These results are to be compared with our numerical data for $d$ below $d_{c}=4$. For $d<4$, we simulated in addition to the CP and sDP also the pair contact process (PCP) introduced by Jensen (see~\cite{Je93} as well as the reviews~\cite{Hi00} and~\cite{Lu04}). The corresponding numerical curves for $U$ are shown in Fig.~(\ref{U_von_d}). From these curves we
find $U (d=3)=0.61$, $U (d=2)=0.704$, and $U (d=1)=0.833$. Our 1-loop
calculation reproduces qualitatively correct the right trend of $U$ as a
function of $d$. Not surprisingly, the quantitative agreement is rather poor
for low dimensions. At least for $d=3$, the $\varepsilon$-expansion estimate is
not too far away from the numerical value. For a field theoretic 1-loop calculation of amplitude ratios, errors of about $10\%$ are typical for $\varepsilon=1$~\cite{BrZi85,ZJ96}, and the deviation of our analytical and numerical results for $\varepsilon=1$ is consistent with that. It is important to note that the value of $U(d)$ for a given dimension is quantitatively the same for the three processes, and that, therefore, $U(d)$  proves to be a true universal signature of the DP class. Moreover, the universality of $U(d)$ for the three processes shows that PCP definitely belongs to the DP universality class.

\section{Finite size effects at  $\bm{d_{\text{c}}}$}
\label{atDc}
Here, we study finite size effects right at $d_{\text{c}}$, where the finite system size is expected to generate logarithmic corrections to the bulk behavior. Guided by lessons learned form previous studies of logarithmic corrections~\cite{JaSt04,Gra04,GHS94,GHS99}, we choose to derive scaling forms in a parametric representation rather than in the more traditional representation featuring nested logarithms. Central to the parametric representation is the parametrization~(\ref{relBetweenVars}) of the system size $L$. Equation~(\ref{abRes}) gives our general parametric results for $a$ and $b$, and Eq.~(\ref{a_log}) specializes the result for $a$ to the critical point. Finally, we compare our parametric result for $U$ to our simulation results.

Past studies of logarithmic corrections in DP~\cite{JaSt04,Gra04} and other systems, e.g., linear polymers~\cite{GHS94,GHS99}, led to the observation that one has to push the analytic calculations beyond the leading logarithmic correction to obtain good agreement between theory and simulations. To go beyond the leading logarithmic correction, we will work in the following, as announced above, to 2-loop order as far as the RGEs are concerned. Concerning the scaling functions, it will still be sufficient, for the most part, to work to 1-loop order. However, here is an important exception: a 1-loop calculation of the scaling function of $a$ does not suffice to determine the next to leading logarithmic correction to $a$ entirely. This subtlety will be discussed as we move along.

Our vantage point for this section will be the general scaling forms for the parameter functions $\hat{k}$, $\hat{\tau}$ and $\hat{g}$ derived in Sec.~\ref{belowDc}, Eqs.~(\ref{parameterFktsScalingFormsGen}) in conjunction with Eqs.~(\ref{parameterFktsScalingFormsIngredients1}). To fill these general scaling forms with live for $d=4$, we must solve the characteristics for this dimension.
In order to make our notation as compact as possible, we will write in the following the RG functions as $f(u)=f_{0}+f_{1}u+f_{2}u^{2}+\cdots$ with $f$ standing ambiguously for $\gamma$, $\zeta$, $\kappa$, and $\beta$. The meaning of the coefficients $f_{0}$, $f_{1}$ and should be evident.

First, we solve the characteristics for $d=4$. The solution to the characteristic for the dimensionless coupling constant $u$, differential equation~(\ref{Char-u}), is given by
\begin{equation}
\ell =\ell (\upsilon)=\ell _{0}\, \upsilon^{-\beta_{3}/\beta_{2}^{2}}\exp\Big[-(\beta_{2}
\upsilon)^{-1}+O(\upsilon)\Big]\,, \label{l(w)}%
\end{equation}
with $\ell _{0}$ being an integration constant. The characteristic~(\ref{charQ}) is readily solved with the result
\begin{equation}
Q(\upsilon)=Q_{0}\, \upsilon^{q_{1}/\beta_{2}}\exp\bigg[\frac{(q_{2}\beta_{2}-q_{1}\beta_{3}%
)}{\beta_{2}^{2}}\upsilon+O(\upsilon^{2})\bigg]\,, \label{Q(w)}%
\end{equation}
with a non universal integration constant $Q_{0}$. 

Next, we choose the flow parameter $\ell$ such that the lattice size $L$ effectively acquires a finite value in the scaling limit:
\begin{equation}
\label{choiceEll}
\ell \, \frac{\mu  L}{2 \pi}=1\, .
\end{equation}
With this choice, $\ell$ and $\upsilon$ tend to zero for $\mu L \to \infty$, and $L$ and $\upsilon$ are related via
\begin{equation}
\label{relBetweenVars}
\left({L}/{L_0}\right) =  \upsilon^{\beta_{3}/\beta_{2}^{2}}\exp\Big[(\beta_{2}
\upsilon)^{-1}+O(\upsilon)\Big]\, ,
\end{equation}
where $L_0 = 2 \pi/(\mu \ell_0)$. Note from this relation that, in contrast to the 1-loop approximation of the RG functions ($\beta_3 \to 0$), the 2-loop approximation leads to an effective $L$-dependence of the nonuniversal length $L_0$ which must not be neglected (see also our discussion of Fig.~\ref{U0d=4} below). Taken together, Eqs.~(\ref{Q(w)}) and (\ref{relBetweenVars}) can be exploited as a parametric representation of the tuple $(L, Q)$ with $\upsilon$ as parameter. This representation has the advantage that the resulting formulas are comparatively compact and, more importantly, that one deals with clean expansion in powers of $\upsilon$.

After this prelude, let us return to the parameter functions. Collecting from the renormalized perturbation calculation results~(\ref{Par3}), the general scaling forms~(\ref{parameterFktsScalingFormsGen}), the solutions of the characteristics~(\ref{Q(w)}), and implementing our choice~(\ref{choiceEll}) of the flow parameter, we obtain
\begin{subequations}
\label{paramFktLogCorr}%
\begin{align}
\hat{k}  &  = \left[ \upsilon^{-1}  -  \frac{3}{4} \sigma^\prime (w) \right]^{-{1}/{6}} \exp \left[ c_{\hat{k}} \upsilon + O \big( \upsilon^2 \big) \right],
\\
\hat{\tau}  &  = \tau \left[ \upsilon^{-1}  -  \frac{3}{4} \sigma^\prime (w) \right]^{-{1}/{3}} \exp \left[ c^{(1)}_{\hat{\tau}} \upsilon + O \big( \upsilon^2 \big) \right]
\nonumber \\
& +  \frac{1}{4} \left(  \frac{2\pi}{L} \right)^2 \upsilon^{{13}/{12}} \left[  \sigma (w) - w \sigma^\prime (w) \right] \exp \left[  c^{(2)}_{\hat{\tau}} \upsilon + O \big( \upsilon^2 \big) \right] ,
\\
\hat{g}  &  = 4 \pi \left[ \upsilon^{-1}  -  \frac{3}{4} \sigma^\prime (w) \right]^{-{2}/{3}} \exp \left[  c_{\hat{g}} \upsilon + O \big( \upsilon^2 \big) \right],
\end{align}
\end{subequations}
where now
\begin{align}
\label{wWithLog}
w=&\, \tau \!\left(  \frac{L}{2\pi} \right)^2 \upsilon^{{1}/{4}} \exp \left[  c_{a} \upsilon + O \big( \upsilon^2 \big) \right]
\nonumber \\
&+ m \frac{L^2}{\pi} \upsilon^{{13}/{12}} \exp \left[ \frac{c_{\hat{k}}}{2}  \upsilon + O \big( \upsilon^2 \big) \right]
\end{align}
and where we have included nonuniversal integration constants stemming from characteristics solutions~(\ref{Q(w)}), viz.\ $X_0$, $X_{\lambda, 0}$ and $X_{\tau, 0}$, in the nonuniversal amplitudes of $\tau$, $h$, and $m$. The coefficients appearing in the exponentials in Eqs.~(\ref{paramFktLogCorr}) and (\ref{wWithLog}) are given by
\begin{subequations}
\label{expCoeffs}
\begin{align}
c_{\hat{k}}  &  =  \frac{\beta_3 \gamma_1  - \beta_2 \gamma_2 }{\beta_2^2} = \frac{25}{1152} + \frac{161}{576} \ln \left(  \frac{4}{3} \right) \approx  0.10211 \, ,
\\
c^{(1)}_{\hat{\tau}}  &  =  \frac{\beta_2 (  \zeta_2 + \kappa_2 -\gamma_2) - \beta_3 ( \zeta_1 + \kappa_1 - \gamma_1) }{\beta_2^2}
\nonumber \\
&= \frac{49}{576} + \frac{53}{288} \ln \left(  \frac{4}{3} \right) \approx  0.13801\, ,
\\
c^{(2)}_{\hat{\tau}}  &  = \frac{\beta_2 (  \zeta_2  -\gamma_2) - \beta_3 ( \zeta_1  - \gamma_1) }{\beta_2^2}
\nonumber \\
&= - \frac{17}{2304} + \frac{263}{1152} \ln \left(  \frac{4}{3} \right)\approx 0.05830 \, ,
\\
c_{\hat{g}} & =  \frac{\beta_2 (  2 \zeta_2  - 3 \gamma_2)  - \beta_3 (2 \zeta_1  - 3 \gamma_1) }{2\beta_2^2}
\nonumber \\
&= \frac{1}{288} + \frac{53}{144} \ln \left(  \frac{4}{3} \right)\approx 0.10936\, ,
\\
c_a & =\frac{\beta_2 \kappa_2 - \beta_3 \kappa_1 }{\beta_2^2}  = \frac{71}{768} - \frac{17}{384} \ln \left(  \frac{4}{3} \right) \approx  0.07971\,  .
\end{align}
\end{subequations}

Now, we revisit $a$ and $b$. Inserting our results~(\ref{paramFktLogCorr}) into definitions~(\ref{abDef}) we find
\begin{subequations}
\label{abRes}
\begin{align}
\label{aRes}
a=&\, \tau \, \frac{L^2}{2\pi} \left[ \upsilon^{-1} -  \frac{3}{4} \sigma^\prime (w) \right]^{{1}/{4}} \exp \left[  c_{a} \upsilon + O \big( \upsilon^2 \big) \right]
\nonumber \\
&+ \frac{\pi}{2}  \left[ \upsilon^{-1} + F(w)  \right]^{-{1}/{2}}  \left[  \sigma (w) - w \sigma^\prime (w) \right] \exp \left[O \big( \upsilon^2 \big) \right]
\\
b= & \, h  \, \frac{L^4}{2\pi} \left[ \upsilon^{-1}  -  \frac{3}{4} \sigma^\prime (w) \right]^{{1}/{2}} \exp \left[   c_{b} \upsilon + O \big( \upsilon^2 \big) \right] ,
\end{align}
\end{subequations}
where
\begin{align}
c_b & =\frac{\beta_2(\gamma_2 - 2 \zeta_2)-\beta_3 (\gamma_1 - 2 \zeta_1)}{2 \beta_2^2} \nonumber\\
&=  \frac{7}{384} - \frac{17}{192} \ln \left(  \frac{4}{3} \right) \approx  - 0.00724 \,  .
\end{align}
At this point, a comment is in order. A full-fledged 2-loop calculation of $a$'s universal scaling function is expected to produce, inter alia, terms of the same order in $\upsilon$ as the 1-loop calculation. Therefore, we have replaced in second line of Eq.~(\ref{aRes}) the 1-loop contribution $\upsilon^{1/2}$ by the bracket containing $F(w)$, where $F(w)$ is a hitherto unknown function. We will leave the calculation of $F(w)$, which will be challenging, to future work.

Finally, let us return to our ratio $U$ of the order parameter moments. As was the case for $d<4$, we are mainly interested in the strong finite size regime $w\ll1$ and, therefore, we approximate $\sigma(w)\approx\sigma(0)+w\sigma^{\prime}(0)$ and $\sigma^{\prime}(w)\approx\sigma^{\prime}(0)$. Focussing on criticality, we set $\tau =0$. The remains $a$, Eq.~(\ref{aRes}), are then
\begin{equation}
\label{a_log}
a = - \frac{4 \ln 2}{\pi}\big[\upsilon^{-1} + K\big]^{-1/2}
\end{equation}
with a universal correction $K = F(0)$. As discussed above, a calculation of $K$ would require to determine the scaling function  of $a$ to 2-loop order. Because corresponding results are currently not at our disposal, we use $K$ as a fitting parameter. Note that $a$ falls off only as $a \sim [\ln (L/L_0)]^{-1/2}$ in $d=4$ compared to the $a \sim (L/L_0)^{-1/2}$ behavior in $d=5$. Thus, it must be expected that $U$ approaches its zero-mode limit $1/2$ even slower for increasing system size than in $d=5$, and that one needs at $d_\text{c}$ even larger systems than above $d_\text{c}$ for the zero-mode theory to provide a good approximation.

Substituting Eq.~(\ref{a_log}) without a further expansion into our scaling function $U_0 (a)$, Eq.~(\ref{U0res}), we obtain our final result for $U_0$ as a function of $\upsilon$. We then use the so-obtained expression for $U$ in conjunction with Eq.~(\ref{relBetweenVars}) as a parametric representation of the tuple $(L, U)$ with $\upsilon$ as parameter, which we plot together with our numerical data in Fig.~\ref{U0d=4}. In the plot, we use $K$ and $L_0$ as fitting parameters. Our best-fit analytical curve (the red middle curve) impressively tracks our data points over the entire range of simulated lattice sizes including sizes as small as $L=4$. As mentioned earlier, the 2-loop RG contribution to Eq.~(\ref{relBetweenVars}) effectively modifies $L_0$. Due to this modification the slope of the continuous curves is significantly reduced for $L<40$ in comparison to the dashed pure 1-loop curve. The introduction of $K$ leads mainly to a rescaling of the nonuniversal length $L_0$ which manifests itself in the modest deviation of the red middle curve from the blue lower curve for $L<10$. Note that up to 2-loop order, one can eliminate $K$ entirely from $U$ via a simple rescaling of $L_0$. Thus, one may view the introduction of a non-zero $K$ as a crude way of accounting for the influence of high loop-orders. Note also, that the effect of the non-zero, fitted $K$ is much smaller than the effective modification of $L_0$ resulting from the 2-loop RG contribution to Eq.~(\ref{relBetweenVars}). Over all, the agreement between theory and simulation is remarkable. This observation reassures us once more about the validity of our analytical and numerical approaches.  Moreover, it underscores the advantages of the parametric representation and makes tangible the necessity of including 2-loop RG results.
\begin{figure}[t]
\includegraphics[width=8.4cm,angle=0]{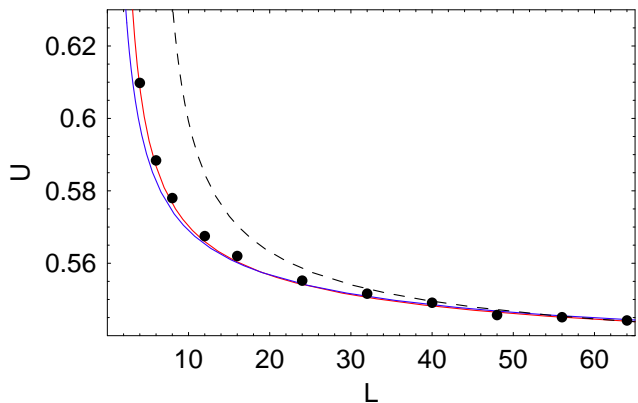}
\caption{(Color online) The universal ratio $U$ versus system size $L$ for $d=4$. The blue lower and the red middle curves represent our analytical results~(\ref{U0res}) and (\ref{a_log}) with $L_0 =2.4,\; K=0$ and $L_0 =1.5,\; K=-0.5$, respectively. For comparison, we included the dashed upper curve, where we have disregarded any 2-loop contributions and where we have fitted $L_0$ to the data points for larger $L$, $L_0 =5.6$. The solid dots stem from our Monte-Carlo simulations of critical sDP on bcc lattices of linear sizes ranging from $L=4$ to $L=64$.}
\label{U0d=4}%
\end{figure}

\section{Concluding remarks}
\label{conclusions}

In summary, we have investigated finite size scaling effects in steady state
systems belonging to the directed percolation universality class. We have
assumed a hypercubic geometry with length $L$, periodic boundary conditions and the presence of an external homogeneous time independent source which prevents the systems to fall into
their absorbing inactive state. We applied a field-theoretic technique based
on an effective response functional (dynamic free energy) for the lowest (homogeneous) mode, which
allowed us to calculate finite size effects within a $1$-loop perturbation
expansion of the higher modes combined with a Markovian approximation. This
latter approximation is indispensable for calculations of strict
non-equilibrium properties of systems without detailed balance. In particular,
it allowed us to calculate the steady state distribution for the lowest mode via
the associated Fokker-Planck equation. Using this distribution, we calculated
explicit scaling forms for the moments of the homogeneous order parameter.
Moreover, we introduced and calculated a ratio $U$ of order parameter moments which allowed us to
analyze universal finite size effects right at the critical point.
Complementary to our analytical work, we performed Monte Carlo simulations
based on the contact process, the site directed percolation process and, on occasion, the pair contact process.

Above and at the upper critical dimension 4, we found remarkable agreement between our analytical and numerical approaches. In these dimensions, the usual coupling constant of the cubic term in the response functional is dangerously irrelevant. Due to this dangerous irrelevance, the universal scaling functions depend on the additional (compared to $d<4$) scaling variable $L/L_0$, where $L_0$ is a nonuniversal length scale. Our results demonstrate that it is necessary to push the diagrammatic calculations beyond 1-loop order to obtain agreement between theory and simulations down to very small systems sizes,  $L/L_0\approx 1$.

For $d$ below 4, we calculated the universal critical values of $U$ in a $\varepsilon$-expansion to order $O (\varepsilon^{3/2})$. The accuracy of this calculation corresponds to that of the calculation of the Binder cumulant of the $\phi^4$-model at the bulk critical point by Brezin and Zinn-Justin~\cite{BrZi85}. 
The agreement between our theory and simulations is within the expectation for a 1-loop calculation that captures terms to $O (\varepsilon^{3/2})$, and, of course, it decreases for decreasing dimensions. However, the universal critical values of $U$ produced by our simulations were identical for all three processes that we simulated and, therefore, $U$ proved to be a true signature of the DP universality class. Moreover, this finding demonstrates that the pair contact process belongs to this class.

On the analytical side, our study is the first investigation, besides the former work of one of us and coworkers, which addresses finite size scaling near absorbing phase transitions. We believe that our approach may be applied to many other non-equilibrium phenomena, and that it can help to improve the understanding of finite-size effects in non-equilibrium systems significantly.

\appendix*

\section{Properties of the functions $D^{(l)}(w)$ and $\sigma (w)$}

Using an exponential representation of the denominators in the sums
(\ref{S_l}), $(r+\mathbf{q}^{2})^{-l}=\Gamma(l)^{-1}\int_{0}^{\infty
}dt\,t^{l-1}\exp(-(r+\mathbf{q}^{2})t)$, we eventually obtain for the
functions $D^{(l)}(w)$ the Laplace-transforms stated in Eq.~(\ref{D_l}).
$D^{(1)}(w)$ and $D^{(2)}(w)$ are smooth functions if $d>4$ (with
$D^{(1)}(0)=4.229$ and $D^{(2)}(0)=21.421$ for $d=5$). Due to the recursion
relation $D^{(l+1)}(w)=\partial D^{(l)}(w)/\partial w$, we can
restrict our attention here to $D^{(1)}(w)$ in order to determine the
remaining properties of $D^{(l)}(w)$ that are used in the main text.

To extract the behavior of $D^{(1)}(w)$ at small arguments, we divide it into
parts,
\begin{equation}
D^{(1)}(w)=I_{1}(w)+I_{2}(w) \label{I-parts}%
\end{equation}
with
\begin{align}
I_{1}(w)  &  =\int_{0}^{\infty}dt\,\mathrm{e}^{-wt}\,\Big(\frac{\pi}
{t}\Big)^{d/2}\Big[1- \sum_{l=0}^{k} \frac{t^{l}}{l!}\,
\mathrm{e}^{-t}\Big]\nonumber\\
& \!\!\! =\pi^{d/2}\Big[\Gamma(1-d/2)w^{d/2-1}
\nonumber\\
&\!\!\!-\sum_{l=0}^{k} \frac{\Gamma(l+1-d/2)}{l!}(1+w)^{d/2-l-1}\Big]\,, \label{I_1}
\end{align}
where $k$ is some integer with $k \geq d/2-1$ to provide integrability at $t=0$. The specifics of the remaining part $I_{2}(w)$ can easily be gathered from Eqs.~(\ref{D_l}) and
(\ref{I_1}). Then, it is straightforward to see $I_{2}(w)$ is an analytic
function of $w.$ Hence, we obtain for small $w=0$ that
\begin{equation}
D^{(1)}(w)=\bar{D}^{(1)}(w)+\pi^{d/2}w^{d/2-1}\left\{
\begin{array}
[c]{ccc}%
\Gamma(1-d/2) &  & \\
\frac{(-1)^{d/2}}{\Gamma(d/2)}\ln w & \text{if} & d/2\in\mathbb{Z}%
\end{array}
\right.  \label{D_l(0)}%
\end{equation}
where $\bar{D}^{(1)}(w)$ is analytic.

To extract the behavior of $D^{(1)}(w)$ for large arguments, we divide this
function in three parts%
\begin{equation}
D^{(1)}(w)=J_{1}(w)+J_{2}(w)+J_{3}(w) \, . \label{J-parts}%
\end{equation}
The behavior of $J_{1}(w)$ and $J_{2}(w)$ for $w\gg1$ is given by
\begin{align}
J_{1}(w)  &  =\int_{\pi}^{\infty}dt\,\mathrm{e}^{-wt}\,\Big[\Big(\frac{\pi}%
{t}\Big)^{d/2}-A(t)^{d}+1\Big]=O\bigl(\mathrm{e}^{-\pi w}%
\bigr)\,,\label{Jot_1}\\
J_{2}(w)  &  =\int_{0}^{\pi}dt\,\mathrm{e}^{-wt}=\frac{1}{w}%
+O(\mathrm{e}^{-\pi w})\,. \label{Jot_2}%
\end{align}
Using the expansion $A(t)^{d}-1=4d\exp(-t)+O\bigl(\exp(-2t)\bigr)$, we find
\begin{align}
J_{3}(w)  &  =\int_{0}^{\pi}dt\,\mathrm{e}^{-wt}\,\Big[\Big(\frac{\pi}%
{t}\Big)^{d/2}-A(t)^{d}\Big]\nonumber\\
&  =\int_{\pi}^{\infty}ds\,\mathrm{e}^{-\pi^{2}w/s}\,\Big(\frac{s}{\pi
}\Big)^{d/2-2}\Big[1-A(s)^{d}\Big]\nonumber\\
&  \approx -4d\int_{\pi}^{\infty}ds\,\,\Big(\frac{s}{\pi}\Big)^{d/2-2}\exp
\Big(-s-\pi^{2}w/s\Big)\nonumber\\
&  \approx -8\pi d\,w^{d/4-1/2}\,K_{d/2}(2\pi\sqrt{w}) \, , \label{Jot_3}%
\end{align}
where $K_{\alpha}(z)$ is the Basset function, for the leading behavior of
$J_{3}(w)$. Using the asymptotic properties of this function, we finally get
\begin{equation}
D^{(1)}(w)=\frac{1}{w}-4\pi d\,w^{(d-3)/4}\,\exp\bigl(-2\pi\sqrt
{w}\bigr)+\ldots\, , \label{D_l(inf)}%
\end{equation}
where the ellipsis denote subleading terms.

For $d\leq4$, we use instead of $D^{(1)}(w)$ the function%
\begin{equation}
\sigma(w)=w(\ln w-1)-\frac{1}{\pi^{2}}D^{(1)}(w)\,,
\end{equation}
where $D^{(1)}(w)$ is taken at $d=4$, and its first derivative $\sigma^{\prime}(w)$ to eliminate the
nonanalytic logarithmic behavior near $w=0$. This function has a power
expansion in $w$ as derived in \cite{JSS88}:%
\begin{equation}
\sigma(w)=\sum_{k=0}^{\infty}\sigma_{k}(-w)^{k}\,
\end{equation}
where
\begin{subequations}
\begin{align}
\sigma_{0} &  =-\frac{8\ln2}{\pi^{2}}\,,\\
\sigma_{1} &= C_{E}+1+\frac{2\ln2}{3}+\frac{6\zeta^{\prime}(2)}{\pi^{2}}\,,
\end{align}
and for $\quad k\geq2$
\begin{equation}
\sigma_{k} =\frac{8(1-1/4^{k})}{\pi^{2}}\zeta(k)\zeta(k+1)\,,
\end{equation}
\end{subequations}
with $C_{E}\approx0.577716$ and $\zeta^{\prime}(2)\approx-0.937548$. For $w\rightarrow\infty$, the function $\sigma(w)$ behaves as
\begin{equation}
\sigma(w)\simeq w(\ln w-1)-\frac{1}{\pi^{2}w}+\frac{16}{\pi}w^{1/4}%
\,\exp\bigl(-2\pi\sqrt{w}\bigr)+\ldots\,,
\end{equation}
up to subleading terms.



\begin{thebibliography}{99}
\bibitem {regions}Here, non-universal amplitudes are included in redefinitions
of $L$ and $r$. The inequalities of Eq.~(\ref{Ber}) reduce for
$d<d_{\mathrm{c}}$ to $1\gg L^{1/\nu}r\gg g_{\ast}^{2}$ where $g_{\ast}%
^{2}\propto\varepsilon+O(\varepsilon^{2})$ denotes the fixed point value of
the coupling constant in the $\varepsilon=d_{\mathrm{c}}-d$ expansion.
\bibitem {footnoteG}A priory, $g$ and $g^{\prime}$ (which, of course must be
positive) are not the same. However, one can identify $g= g^{\prime}$ after a
straightforward rescaling of $s$, $h$, and $\zeta$.
\end{thebibliography}
\end{document}